\documentclass[12pt]{article}

\usepackage[margin=1in]{geometry}
\usepackage{setspace}
\usepackage{lineno}
\usepackage{graphicx}
\usepackage{caption}
\usepackage{amsmath, amssymb}
\usepackage{times}
\usepackage{hyperref}
\usepackage[numbers,sort&compress]{natbib}
\usepackage{booktabs}
\usepackage{siunitx}

\title{\textbf{Reduced cloud cover errors in a hybrid AI-climate model through equation discovery and automatic tuning}}
\author{
Arthur Grundner$^{1,*}$, Tom Beucler$^{2,3}$, Julien Savre$^{1}$, Axel Lauer$^{1}$, \\ Manuel Schlund$^{1}$,  Veronika Eyring$^{1,4}$ \\
\\
\small $^1$Deutsches Zentrum für Luft‐ und Raumfahrt e.V. (DLR), Institut für Physik der Atmosphäre, \\ \small Oberpfaffenhofen, Germany \\
\small $^2$Faculty of Geosciences and Environment, University of Lausanne, Lausanne, Switzerland \\
\small $^3$Expertise Center for Climate Extremes, University of Lausanne, Lausanne, Switzerland \\
\small $^4$University of Bremen, Institute of Environmental Physics (IUP), Bremen, Germany \\
\small $^*$Corresponding author: arthur.grundner@dlr.de
}
\date{}

\begin{document}

\maketitle

% Abstract
\begin{abstract}
\noindent
Cloud-related parameterizations remain a leading source of uncertainty in climate projections. Although machine learning holds promise for Earth system models (ESMs), many data-driven parameterizations lack interpretability, physical consistency, and smooth integration into ESMs. Here, a two-step method is presented to improve a climate model with data-driven parameterizations. First, we incorporate a physically consistent cloud cover parameterization—derived from storm-resolving simulations via symbolic regression, preserving interpretability while enhancing accuracy—into the ICON global atmospheric model. Second, we apply the gradient‑free Nelder–Mead optimizer to automatically recalibrate the hybrid model against Earth observations, tuning in nested stages (2‑, 7‑, 30- and 365‑day runs) to ensure stability and tractability. The tuned hybrid model substantially reduces long-standing biases in cloud cover—particularly over the Southern Ocean (by 75\%) and subtropical stratocumulus regions (by 44\%)—and remains robust under +4K surface warming. These results demonstrate that interpretable machine-learned parameterizations, paired with practical tuning, can efficiently and transparently strengthen ESM fidelity.
\end{abstract}

\vspace{1cm}

% Main text
\section*{Introduction}
When numerically modeling the dynamics of a physical field, it needs to be overlaid by a discrete grid. Processes that are happening inside individual grid cells and thus cannot be resolved may have a significant impact on the overall dynamics and must therefore be inferred from the available grid-scale information. In climate models, so-called parameterizations tackle this task \cite{stensrud2009}. It is natural to learn from data where (e.g., process) knowledge is lacking, therefore, the development of data-driven and machine learning based parameterizations has become widespread in recent years \cite{eyring2024pushing, eyring2024pAIempowered, gentine2018could, grundner2022deep, heuer2024interpretable, o2018using}. Because of their representation power, neural networks are a common choice for data-driven modeling \cite{rasp2018deep}, despite coming with some clear and significant drawbacks: They often deteriorate the interpretability of climate models, making it difficult to demonstrate their physical consistency, they degrade the climate model's applicability to climate change scenarios \cite{reichstein2019deep}, and they increase the computational burden when replacing a parameterization that used to be formulated as a simple equation \cite{grundner2024data}. 

% Very short summary of paper structure.
In the development of all types of data-driven parameterizations, the training and testing of a scheme is performed either `offline' (i.e., in a stand-alone fashion on a reference dataset) or `online' (i.e., implemented in and interacting with the dynamics of a climate model) \cite{grundner2022deep, bracco2025}. Online training requires either a differentiable host climate model \cite{frezat2022posteriori} or costly inverse methods \cite{christopoulos2024}, making offline training the dominant approach for data-driven parameterizations. However, these offline-trained parameterizations have yet to demonstrate consistent performance online, due to the need for a tuning step to adjust the climate model to its new parameterization \cite{chen2023neural, morcrette2024scale}. Such a tuning step usually requires substantial time and expert knowledge. Its execution also becomes less straightforward as deep learning-based parameterizations no longer include such physics-related tunable parameters. Thus, many of the offline-trained models are never tested online, leading to increased doubts about their usefulness for climate modeling \cite{frezat2022posteriori}. After all, it is not obvious that a data-driven parameterization can overcome this `optimization dichotomy' \cite{rasp2021optimization} of being trained in one setting while expected to perform well in a different setting.

In this paper we do not only demonstrate the strong online skill of our offline-trained, data-driven parameterization, but also introduce a new automatic tuning pipeline that enables these results.
More specifically, we consider an \textit{analytical equation} (equation (\ref{data_driven_eq})) from \cite{grundner2024} that has been trained offline to parameterize cloud fraction, reducing related systematic model errors towards improved climate projections. The equation was derived from a high-resolution simulation using a symbolic regression library \cite{cranmer2023interpretable} for equation discovery \cite{huntingford2025potential, lai2024machine}. It is the product of a data-driven approach that provides accuracy and yet retains the interpretability, physical consistency and computational efficiency of the traditional parameterization. However, its application in a climate model has not yet been demonstrated. To do so, we first implement the data-driven cloud cover equation into ICON-A \cite{giorgetta2018}, the atmospheric component of the ICOsahedral Non-hydrostatic (ICON) climate model. We then set up an automatic tuning procedure following a multi-objective approach \cite{neelin2010} to tune the model, hereafter called `ICON-A-MLe' where MLe stands for Machine Learning enhanced. The basic idea is to tune as much as possible using simulations that are as short as possible. Here, we build on the finding that biases resulting from fast physical processes can already be identified in short simulations of a day, week or month \cite{xie2012}. Compared to other recent automated tuning methods for climate models (e.g., \cite{bonnet2024, roach2018}), our tuning procedure is simpler and thus more efficient and better generalizable.

We begin by outlining the tuning pipeline and its impact on ICON model results. We then compare the tuned ICON-A-MLe model with observations, as well as to models participating in the Coupled Model Intercomparison Project Phase 6 (CMIP6 \cite{eyring2016overview}) and to an automatically and manually tuned version of ICON-A. For the evaluation, we perform a historical Atmospheric Model Intercomparison Project (AMIP) simulation from 1979-1999, which has prescribed sea surface temperatures (SSTs) and sea ice concentrations, and evaluate it with Earth observations to assess its accuracy. For this we use the Earth System Model Evaluation Tool (ESMValTool), an established community diagnostic and performance metrics tool for the evaluation and analysis of Earth system models \cite{righi2020,eyring2020,lauer2020,weigel2021,esmvalcore,esmvaltool,lauer2025monitoring}, recently extended to be able to process ICON output without model postprocessing \cite{schlund2023}. To isolate the impact of the new scheme from the tuning, we also present a ``scheme-swap'' experiment with subsequent retuning of the cloud cover schemes. Following this, we provide a physical analysis linking the new scheme's analytical terms to the improved cloud representation.
Finally, we demonstrate its robustness by applying ICON-A-MLe to a significantly warmer climate.

\section*{Overview of the Automatic Tuning Pipeline}

\begin{figure*}[tbhp]
\centering
\hspace*{-2em}
\captionsetup{margin=0.5cm} % To confine the caption to the image
\includegraphics[width=18cm]{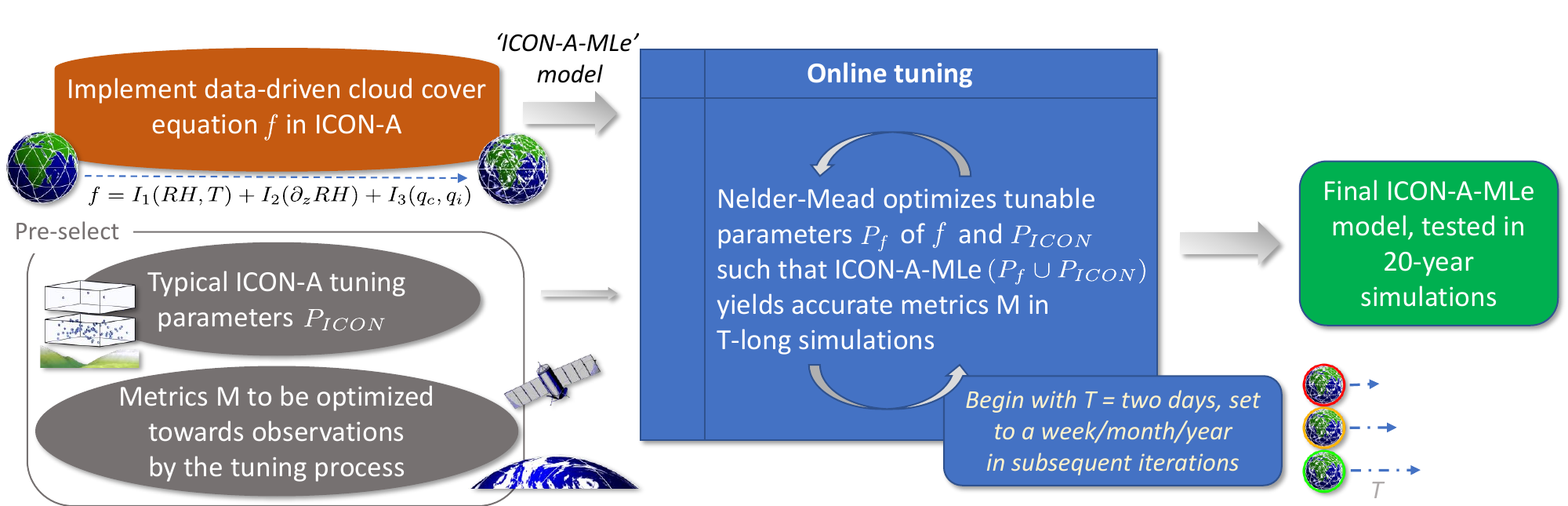}
\caption{The automatic tuning pipeline of ICON-A-MLe, i.e., the ICON atmospheric climate model including our data-driven cloud cover equation (equation (\ref{data_driven_eq})). The only manual step involves selecting the climate metrics to be optimized, such as globally averaged cloud cover, and determining which tunable parameters should be adjusted. Reasonable target ranges for these metrics are derived from observational and reanalysis data. The core of the tuning pipeline consists of the Nelder-Mead algorithm iteratively searching for a setting of the tunable parameters that enables ICON-A-MLe simulations to optimize all specified metrics. The length of these simulations and the strictness of evaluating the metrics increases with the number of iterations. The underlying strategy leverages the fact that some climate metrics exhibit rapid responses to parameter changes, allowing for efficient feedback and early adjustments. Finally a historical AMIP simulation is benchmarked to Earth observations with the ESMValTool}
\label{fig:pipeline_sketch}
\end{figure*}

To effectively integrate the data-driven cloud cover equation (equation (\ref{data_driven_eq})) into ICON-A, we developed a simple and efficient automatic tuning pipeline (Fig. \ref{fig:pipeline_sketch}). This pipeline uses the Nelder–Mead optimization algorithm to calibrate a set of 24 physical and empirical parameters—spanning cloud microphysics, convection, and the new parameterization (Table \ref{tab_si:tab_S3})—against observationally derived target metrics (Methods).

Utilizing computational resources carefully, the tuning process proceeds in stages, beginning with short 2-day simulations and increasing in duration to 7-day, 30-day, and finally 365-day simulations. At each stage, simulation outputs are compared to observational constraints on key global metrics such as the top-of-the-atmosphere (TOA) radiation balance, shortwave (SW)/longwave (LW) cloud radiative effects (CRE), precipitation, and cloud water paths. The objective function guiding optimization quantifies the normalized distance between simulated values and observational ranges. The shorter the simulations in a given stage, the more those ranges are relaxed informed by the typical variability of a given metric in a reference ICON simulation. Each optimization stage refines the parameter space based on the best result from the previous stage. The final step includes an extrapolated parameter update to balance short- and long-term simulation objectives, followed by a small number of year-long simulations to identify an optimal parameter set. To minimize the objective function within each stage, we use the Nelder-Mead algorithm \cite{nelder1965simplex}; a robust, derivative-free method that moves and tumbles a ``simplex'' (which is a triangle in a 2-dimensional parameter space, a tetrahedron in 3D, and so on) through the parameter space, iteratively discarding the worst-performing vertices. We apply it in this study to automatically find the set of model parameters that minimizes the deviation between our simulation output and observational data. This iterative strategy allows for robust and efficient convergence toward physically realistic climate simulations.

\begin{figure}[tbhp]
\centering
\includegraphics[width=15cm]{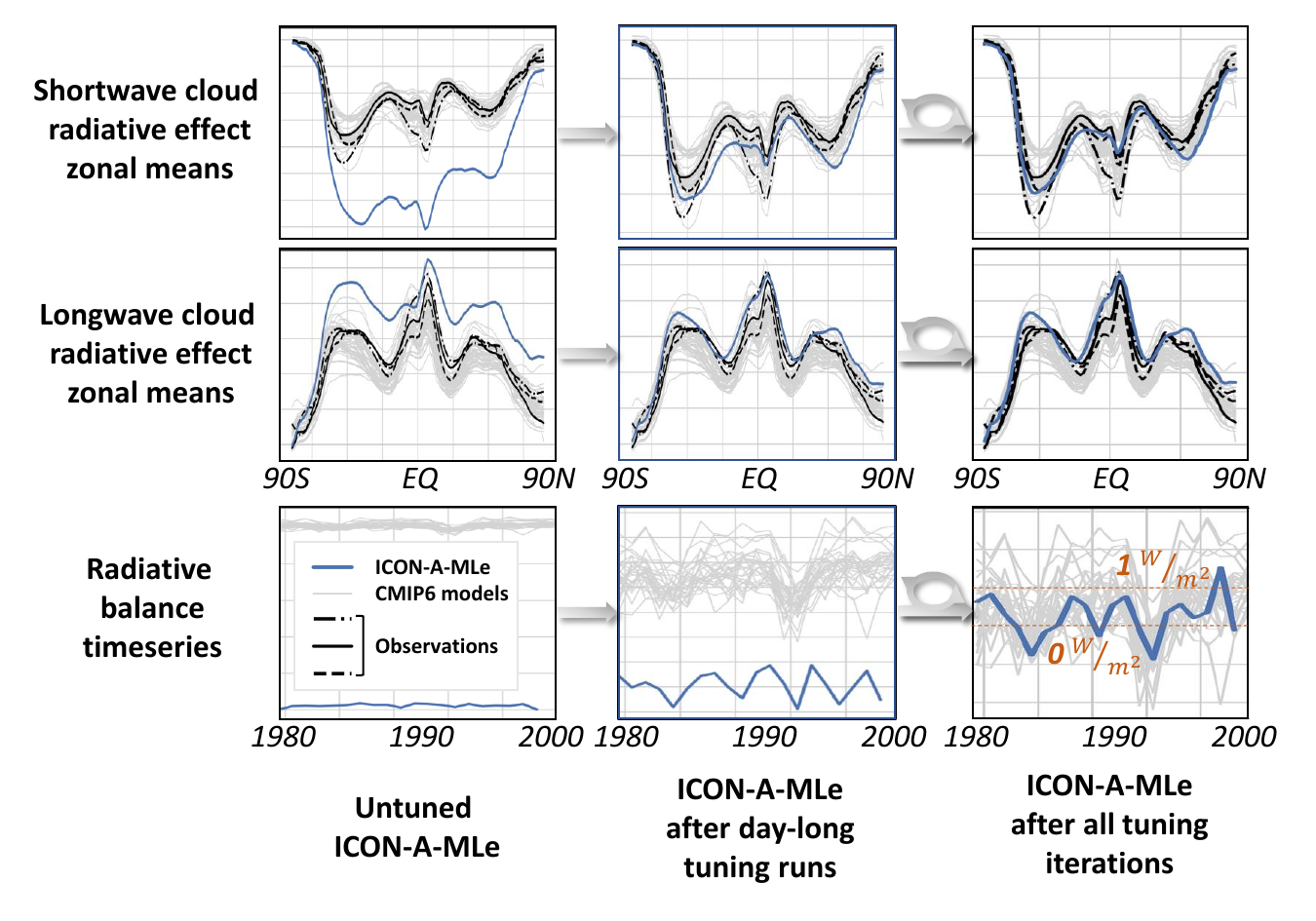}
\caption{A qualitative evaluation of 20-year ICON-A-MLe simulations using parameter settings extracted at three different stages of the tuning pipeline. The panels display radiative measures computed at the top of the atmosphere, illustrating the model's progression through the tuning process. Circular arrows denote intermediate tuning steps involving week- and month-long ICON-A-MLe simulations. Observational references are derived from MERRA2 \cite{gelaro2017modern}, CERES \cite{loeb2018clouds}, and ISCCP \cite{zhang2023global}. The solid gray lines represent historical CMIP6 model simulations \cite{eyring2016overview}, providing a benchmark for comparison}
\label{fig:pipeline_eval}
\end{figure}

Figure \ref{fig:pipeline_eval} illustrates the impact of the automatic tuning pipeline on the performance of ICON-A-MLe in 20-year simulations for three key radiative metrics (zonal means of the absorbed SW and outgoing LW radiation are provided in Fig. \ref{fig_si:fig_S6}). When the data-driven equation (equation (\ref{data_driven_eq})) with its original parameters (Table \ref{tab_si:tab_S3}) is first implemented in ICON, these metrics deviate significantly from observations without tuning. After each optimization stage, these metrics gradually improve with the top-of-the-atmosphere radiative balance assuming realistic values at the end of the tuning process.

\section*{Evaluation of 20-year ICON Simulations}
In this section, we compare the tuned ICON-A-MLe and ICON-A models with observational reference datasets, focusing on the mean state of the simulated climate. Specifically, we conduct a quantitative evaluation of the 20-year historical AMIP simulations referenced in the green box of Fig. \ref{fig:pipeline_sketch}.

\begin{figure*}[tbhp]
\hspace*{-3em}
\includegraphics[width=19cm,height=11cm]{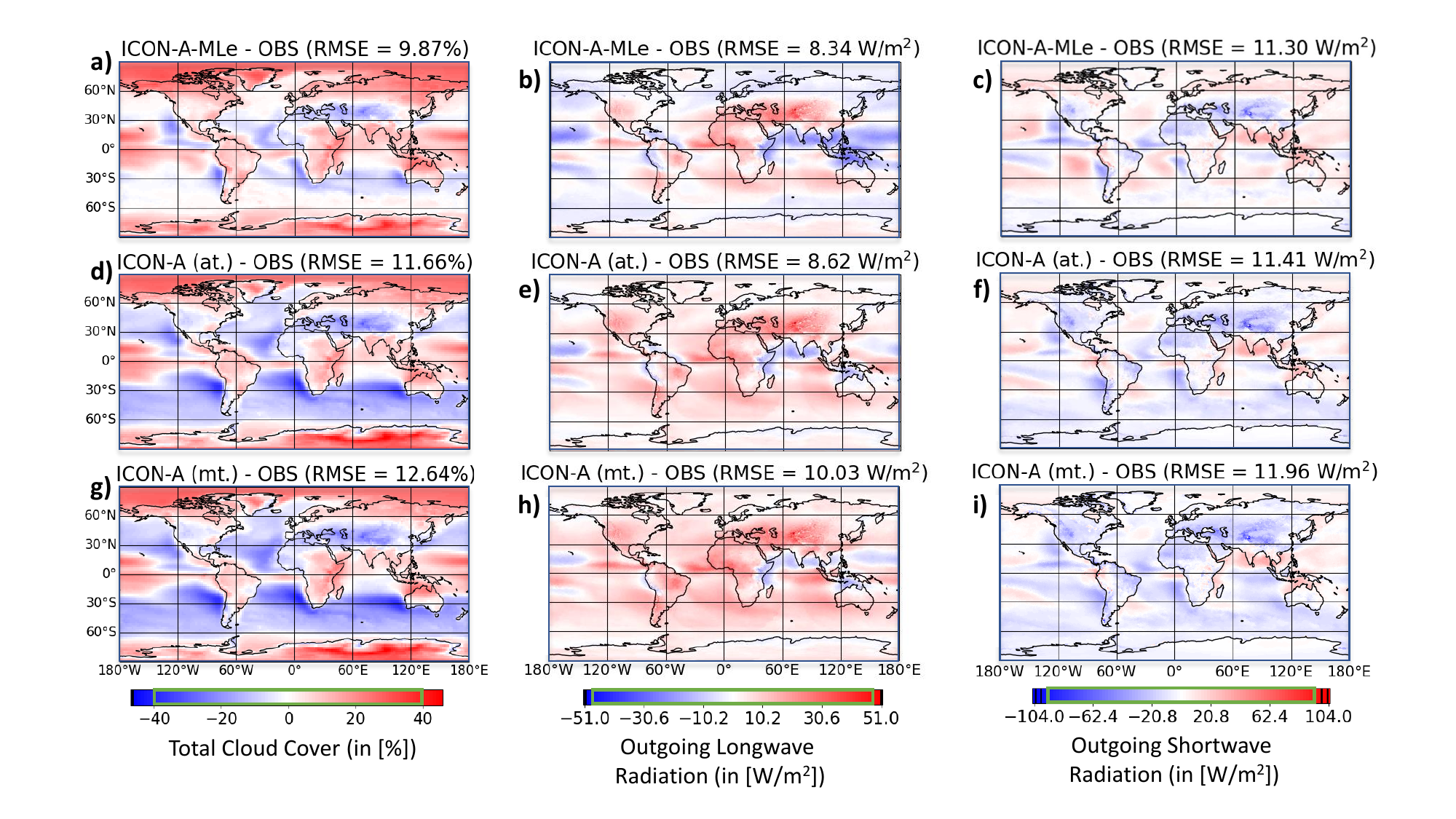}
\caption{Biases in three key climate metrics, temporally averaged over 20-year simulations (1979–1999), for our automatically tuned (at.) ICON-A-MLe (first row) and ICON-A models (second row), and a manually tuned (mt.) ICON-A reference (third row). Root-mean-square errors (RMSEs) are computed after remapping the data onto a horizontal grid with nearly identical grid cell sizes. While the ICON-A panels utilize the full range of each colorbar (otherwise indicated by black vertical lines in the colorbars), the values from the ICON-A-MLe model are confined to the green rectangles. To enhance the robustness of our observational reference (OBS), we take an average across multiple observational cloud cover datasets (CLARA-AVHRR \cite{CLARA_AVHRR}, ESA CCI \cite{stengel2017cloud}, MODIS \cite{platnick2003modis}, PATMOS \cite{heidinger2014pathfinder}), as well as two reanalysis products (MERRA2 \cite{gelaro2017modern}, ERA5 \cite{hersbach2020era5}). For radiative metrics, we use ESA CCI \cite{stengel2017cloud} (1982–1999) and ISCCP \cite{zhang2023global} (1984–1999), as these observations are among the most reliable spanning the majority of the simulated 1979–1999 period. The plots in the panels were created by visualizing the output from ESMValTool (v2.12.0, \url{https://www.esmvaltool.org/}) using the Psyplot software (v1.5.1, \url{https://psyplot.github.io/}) \cite{psyplot} for final customization}
\label{fig:iconml_eval}
\end{figure*}

% Evaluation 

% OBS_ESACCI-CLOUD_sat_AVHRR-AMPM-fv3.0_Amon_rlut_198201-199912
% OBS_ESACCI-CLOUD_sat_AVHRR-AMPM-fv3.0_Amon_rsut_198201-199912
% CLARA-AVHRR (19990101/20181231), ERA5 (20020101/20211231), ESACCI-CLOUD (19970101/20161231), MERRA2 (20020101/20211231), MODIS (20030101/20181231), PATMOS-x (19970101/20161231)

% Concerning Figure 3 (ICON-A-ML vs ICON-A)
% Why those three metrics (total cloud cover, SW/LW radiation)?
% Are the ICON-A-ML results much better than those of ICON-A?

Total cloud cover is the first metric to evaluate when testing a new cloud cover parameterization. Its importance cannot be overstated: clouds are fundamental to the Earth’s energy budget, as they can either cool the Earth by reflecting sunlight or warm it by trapping heat. Their role in modulating precipitation and humidity further underscores their influence on local and regional climates \cite{zelinka2020}.

The first column of Fig. \ref{fig:iconml_eval} demonstrates that ICON-A-MLe achieves a notably more accurate representation of the geographical distribution of temporally averaged total cloud cover compared to the manually and automatically tuned ICON-A setups (global RMSE reduced by 15.4\% from Panel d) to a)). These improvements are most evident in maritime regions, particularly in regions dominated by marine low-level clouds \cite{muhlbauer2014climatology}; the Southern Ocean (RMSE reduced by 75.8\%), the west coasts of Chile/Peru (by 49.9\%), Namibia/Angola (by 54.2\%), California (by 24.1\%), Morocco (by 41.8\%), and Australia (by 53.3\%). These are precisely the regions that exhibit the largest positive differences between Panels a) and d) (see Fig. \ref{fig_si:fig_S7} a)). 
%Over land and over the poles the biases in cloudiness are very similar. 
Polewards of 60$^{\circ}$N and 60$^{\circ}$S, ICON-A significantly overestimates cloud cover. While ICON-A-MLe shows an increased cloudiness over Canada and Greenland even further (locally by up to 23\%) compared to ICON-A, it displays reduced cloud cover over most of Antarctica (by nearly 10\%), thereby decreasing the corresponding bias to below 40\% cloud cover. As the remaining deviation cannot be entirely explained by the substantial measurement uncertainty in polar regions (up to 15\% \cite{liu2022analysis}), this motivates further refinement in future studies.

While the cloud cover root-mean-square error (RMSE) is lower for the automatically tuned (Panel d)) compared to the manually tuned (Panel g)) ICON-A simulation ($\Delta$RMSE = $-0.98\, $W/m$^2$), their bias plots are almost indistinguishable as regional differences are typically rather small ($\leq$ 10\% cloud cover as can be seen in Fig. \ref{fig_si:fig_S7} d)).

Improvements we find in cloud cover are most valuable if, to some extent, they translate to improved radiative metrics. After all, an accurate representation of LW and SW radiation is crucial for ensuring that the climate model can effectively simulate the Earth's energy budget—a fundamental component for projecting temperature and other climate-relevant parameters. The second column of Fig. \ref{fig:iconml_eval} reveals that the manually tuned ICON-A (Panel h)) generally overestimates outgoing LW radiation, particularly over mountainous regions. This overestimation arises from the model’s coarse-resolution grid, which smooths topography and causes mountains appear lower—and therefore warmer—than in reality. The automatically tuned ICON-A (Panel e)) shows a reduced outgoing LW radiation globally, except for a localized increase (up to 15 W/m$^2$) around the Philippines (see also Fig. \ref{fig_si:fig_S7} e)). ICON-A-MLe further reduces outgoing LW radiation, bringing it closer (with exceptions over the tropics) to observations. 
Especially over oceans, the increase in cloud cover causes reduced LW radiation now being emitted from cloud tops that are cooler than the ocean's surface. However, ICON-A-MLe tends to underestimate outgoing LW radiation over the tropical western Pacific. This bias stems partly from a (more accurate) increase in cloud ice (see Fig. \ref{fig_si:fig_S8}), which forms at higher altitudes and thus emits less LW radiation.

The difference between the SW radiation maps (third column of Fig. \ref{fig:iconml_eval}) from ICON-A-MLe (Panel c)) and the automatically tuned ICON-A (Panel f)) are highly correlated with those of cloud cover (see also Fig. \ref{fig_si:fig_S7}). The largest discrepancies occur in regions with an abundance of marine low-level clouds, where increased cloud coverage leads to more SW radiation being reflected back to space. The increase in cloudiness turns a slight underestimation of the reflected TOA SW radiation into a slight overestimation and results in a small global RMSE improvement ($\Delta$RMSE = $-0.11\, $W/m$^2$). Meanwhile, the automatically tuned ICON-A’s SW radiation map itself shows improvements ($\Delta$RMSE = $-0.55\, $W/m$^2$) over the manually tuned version due to its increased marine low-level cloud coverage and consequently increased reflected SW radiation.
% Generally better RMSEs and inf norm.
Across all three metrics—cloud cover, outgoing LW, and reflected SW radiation—ICON-A-MLe consistently achieves the lowest RMSE values, while the manually tuned ICON-A model performs the worst. This underscores the improvement that can be achieved through the combination of the automatic tuning procedure and the data-driven cloud cover equation. % The vertical lines in the colorbar provide further evidence: the maximum cloud cover cell discrepancy for ICON-A-MLe is significantly lower (39.8\%) compared to the ICON-A simulations (46.6\% and 46.3\%). 

% What about the other metrics? Refer to Table in SI.
In addition to the three metrics presented in Fig. \ref{fig:iconml_eval}, Table \ref{tab_si:tab_S1} provides a comprehensive comparison of the biases (RMSE and R$^2$-values) for other key climate variables across the three simulations. While minor increases in RMSE can occur, the R$^2$-values of ICON-A-MLe are generally on par with or better than those of the ICON-A models. Exceptions are the simulation of the LW cloud radiative effect (LWCRE), which is better in the auto-tuned ICON-A model ($\Delta R^2 = -0.14$), and the ice water path, which is slightly better in the manually tuned ICON-A model ($\Delta R^2 = -0.02$). The most striking improvement ($\Delta R^2 = 0.22$) is observed in the liquid water path, defined as the vertically integrated amount of cloud water. 
% This improvement is unsurprising given ICON-A-MLe's enhanced ability to capture marine low-level clouds, which are primarily composed of liquid water droplets.

\section*{Isolating the Added Value of the Data-Driven Cloud Cover Scheme}

While the differences between the baseline ICON-A and the ICON-A-MLe models (Fig. \ref{fig:iconml_eval}), both fully tuned in the same automatic manner, provide indications that the data-driven cloud cover scheme itself contributes to the error reduction, in this section, we quantify more precisely how much of the improvements gains of the previous section we can solely attribute to the data-driven scheme (as opposed to the automatic tuning procedure). To do this, we designed two configurations to test each scheme in an environment optimized for the \textit{other} scheme: \\

\textbf{ICON-A-MLe*}: We start with the fully tuned $\text{ICON-A}$ configuration (the baseline model's optimized environment). We then replace its native cloud scheme with our new data-driven scheme and retune only the parameters of the new scheme. This tests the new scheme in a potentially difficult environment as the tuning procedure is not allowed to mitigate other potential biases present in the climate model.

$\textbf{ICON-A*}$: Conversely, we start with the fully tuned $\text{ICON-A-MLe}$ configuration (the new scheme's optimized environment). We replace the data-driven scheme with the native scheme and retune only the parameters of the native scheme. This tests the native scheme in an environment optimized for the data-driven scheme. \\

\begin{figure*}
\centering
\hspace*{-3em}
\captionsetup{margin=0.2cm} % To confine the caption to the image
\includegraphics[width=19cm]{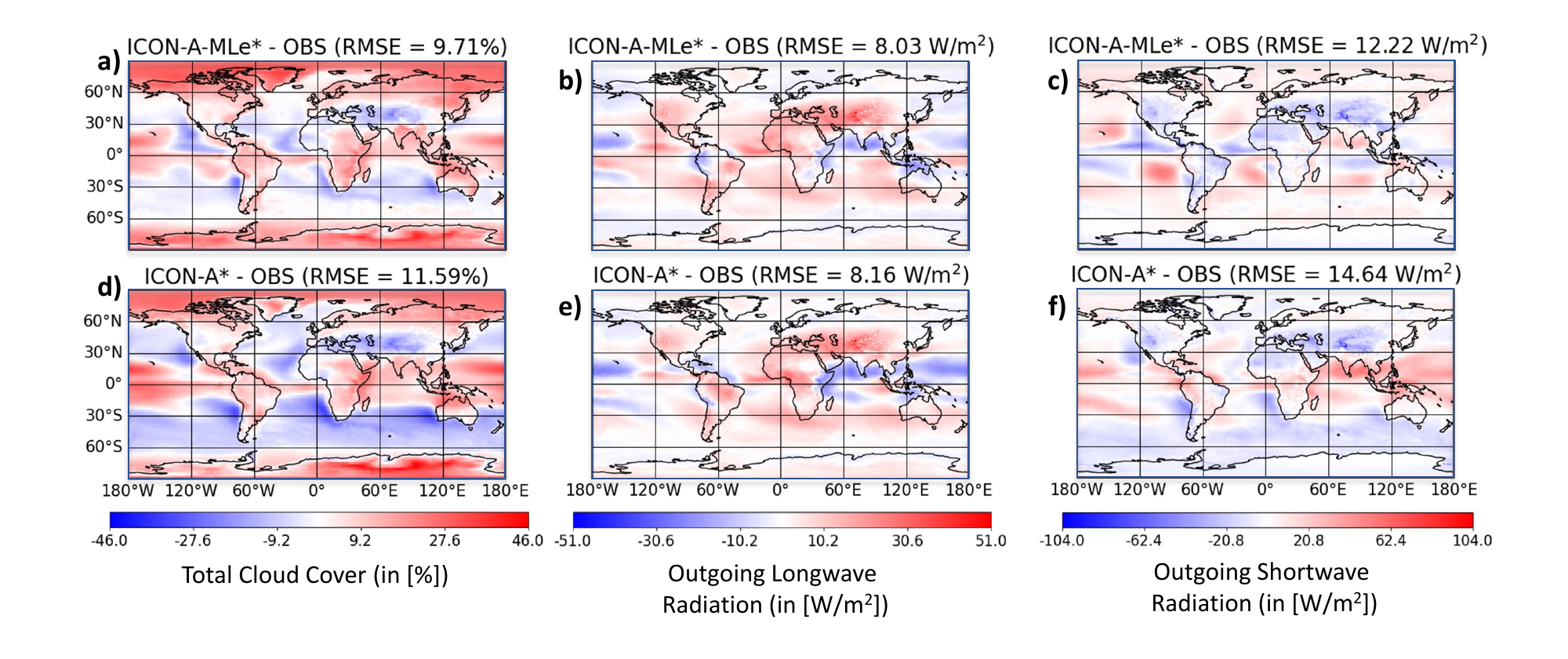}
\caption{Evaluation of ICON-A-MLe* and ICON-A* 20-year simulations as in Fig. \ref{fig:iconml_eval}}
\label{fig:iconstar}
\end{figure*}

In Fig. \ref{fig:iconstar} we evaluate these two configurations by showing the same statistics as in Fig. \ref{fig:iconml_eval}. The spatial bias patterns remain consistent, confirming the signature behavior of each scheme regardless of the tuning environment. For instance, the ICON-A-MLe models tend to overestimate SW radiation over the Eastern Pacific and the ICON-A models consistently show a strong total cloud cover bias over the Southern Ocean. As opposed to their counterparts, some models seem to trade in more distinct regional biases for reduced global ones (e.g., LW radiation over the Pacific in ICON-A* or ICON-A-MLe). \\
Notably, the data-driven scheme performs very well in the $\text{ICON-A}$ environment ($\text{ICON-A-MLe*}$); its global RMSE for total cloud cover (9.71\%) and LW radiation (8.03 $\text{Wm}^{-2}$) is even slightly lower than in the fully tuned $\text{ICON-A-MLe}$ simulation in which all parameters were tuned together in a pursuit to reach a highly dimensional optimum. In contrast, the native scheme struggles significantly in the environment tuned for the data-driven scheme ($\text{ICON-A*}$), with its SW radiation RMSE increasing to 14.64 $\text{Wm}^{-2}$.

We can quantify the average impact of the data-driven scheme by averaging its performance gain across both experimental setups. We calculate the average change in RMSE by comparing the new scheme to the old scheme within each of their respective ``background" environments:

\begin{align*}
    \text{Avg. } \Delta\text{RMSE} &= \frac{1}{2} \Big( \big( \text{RMSE}(\text{ICON-A-MLe*}) - \text{RMSE}(\text{ICON-A}) \big) \\&+ \big( \text{RMSE}(\text{ICON-A-MLe}) - \text{RMSE}(\text{ICON-A*}) \big) \Big)
\end{align*}

For the metrics of Fig. \ref{fig:iconml_eval} this calculation yields
\begin{align*}
    \begin{cases}
        (9.71 - 11.66 + 9.87 - 11.59 )/2\% = \mathbf{-1.84}\%, \, \, \, &\text{total cloud cover,}  \\
        (8.03 - 8.62 + 8.34 - 8.16 )/2\,\text{Wm}^{-2} = \mathbf{-0.21}\,\text{Wm}^{-2}, \, \, \, &\text{longwave radiation,}  \\
        (12.22 - 11.41 + 11.3 - 14.64 )/2\,\text{Wm}^{-2} = \mathbf{-1.27}\,\text{Wm}^{-2}, \, \, \, &\text{shortwave radiation.}  \\
    \end{cases}
\end{align*}

The negative values for all three metrics confirm that, on average across both tuned environments, the data-driven cloud cover scheme robustly reduces the RMSE compared to the native scheme.

This finding is supported by Table \ref{tab_si:tab_S2}, which shows that the data-driven scheme generally improves simulation results across all considered metrics. The most pronounced positive impacts are on liquid water path (LWP), which sees an $R^2$-score improvement of $\approx 0.5$, and total cloud cover, which benefits from an RMSE reduction of $1.77\%$. Overall, $R^2$-scores improve for five metrics and remain neutral for three. The only metric that degrades is the LWCRE, with an $R^2$ decrease of $0.09$. This appears to be a necessary trade-off, as the baseline model's high $R^2$-score for LWCRE was achieved at the cost of underestimating both LWP and total cloud cover. \\

\section*{Physical Drivers of Improved Cloud Cover in ICON-A-MLe} 
\begin{figure*}[!htb]
\centering
% \hspace*{-3em}
\captionsetup{margin=0.2cm} % To confine the caption to the image
\includegraphics[width=17cm]{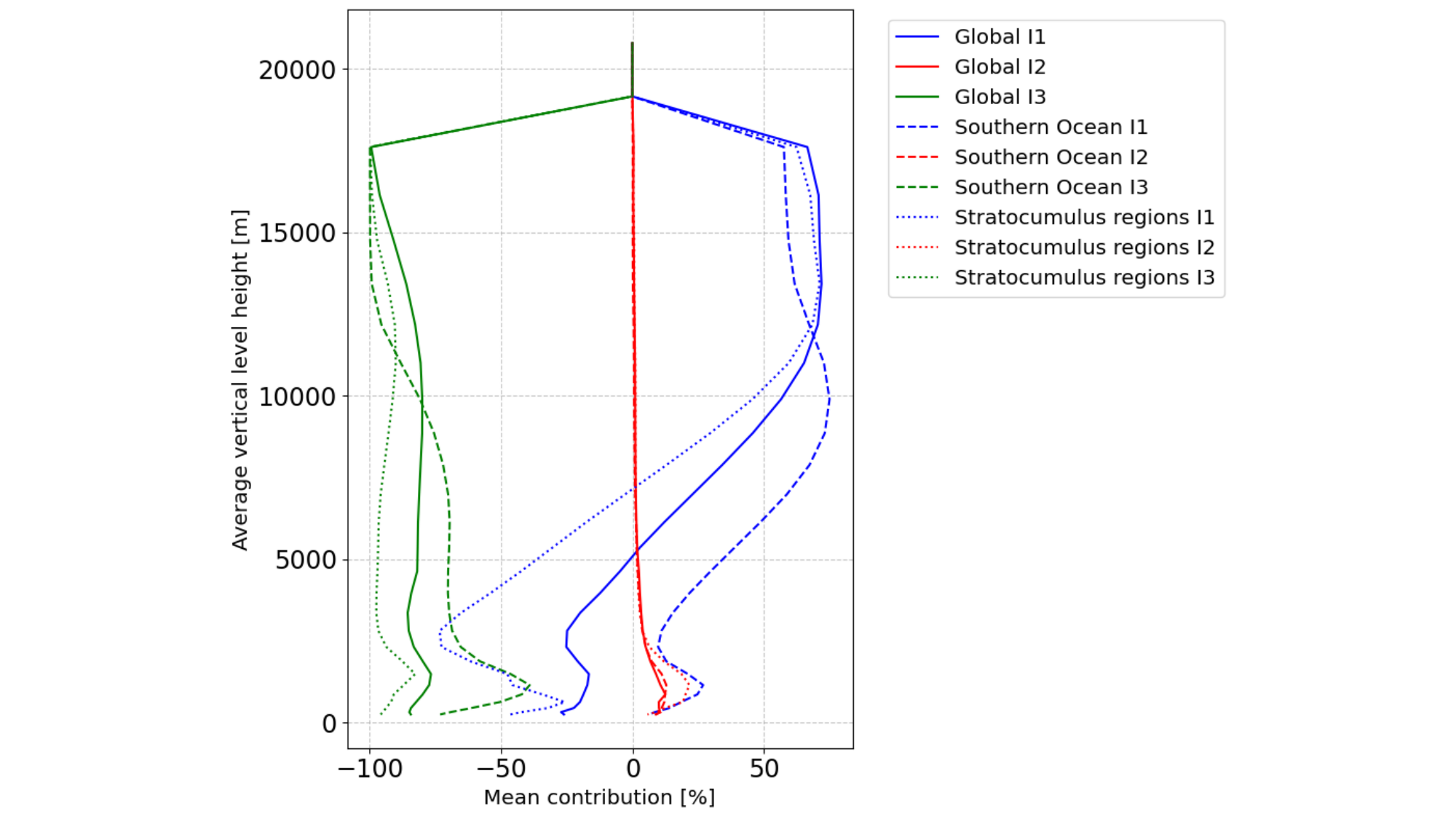}
\caption{Vertical profiles of the contributions from each term ($I_1$, $I_2$, $I_3$) of the data-driven cloud cover equation (\ref{data_driven_eq}) to the total cloud fraction on each height level. Values are annual means from an $\text{ICON-A-MLe}$ simulation (1979–1980), spatially averaged globally, as well as over two distinct regions. The 'Stratocumulus' regions combine the west coasts of Chile/Peru, Namibia/Angola, California, Morocco, and Australia}
\label{fig:fig_ix_contr}
\end{figure*}

A primary motivation for developing the new analytical parameterization was its physical interpretability \cite{grundner2024}. The superior performance of the ICON-A-MLe configuration in maritime regions is not just a product of the optimization methodology but  directly linked to its physical formulation (equation \ref{data_driven_eq}).  Figure \ref{fig:fig_ix_contr} confirms that the improvements in specific regions correspond directly to the intended physical function of these terms. \\

\textbf{Southern Ocean Cloudiness (The $I_1$ Term)} \\
The most significant reduction in cloud cover biases over the Southern Ocean can be attributed to the $I_1$ term. As detailed in \cite{grundner2024}, this term introduces a more nuanced dependence on temperature ($T$) than the baseline Sundqvist scheme. This is crucial for improving the representation of low-level clouds in this region, which the baseline model struggled to capture. The tuning process optimized the parameters governing $I_1$, allowing it to become most active in this specific regime and leading to the documented performance gains. \\

\textbf{Subtropical Stratocumulus Decks (The $I_2$ Term)} \\
In contrast, the improved simulation of subtropical stratocumulus decks is driven by the $I_2$ term. This term is formulated to track the vertical gradient of relative humidity, $\partial_z\mathrm{RH}$. It is therefore most sensitive to the strong, low-level temperature inversions that cap these marine boundary layer clouds and are accompanied by a rapid decrease in relative humidity with height. The activation of the $I_2$ term in these specific regions allows the model to better represent the sharp cloud tops and maintain the stratocumulus layers, correcting known biases in the baseline model. \\

In summary, the results of the tuning validate the physical hypothesis behind the parameterization's design. The $I_1$ and $I_2$ terms function as "dials" that the tuning methodology can adjust, but their design ensures that these adjustments correspond to distinct and physically meaningful processes in specific climate regimes.

\section*{Application of ICON-A-MLe to Warmer Climates}
\begin{figure*}[!htb]
\centering
\hspace*{-3em}
\captionsetup{margin=0.2cm} % To confine the caption to the image
\includegraphics[width=19cm]{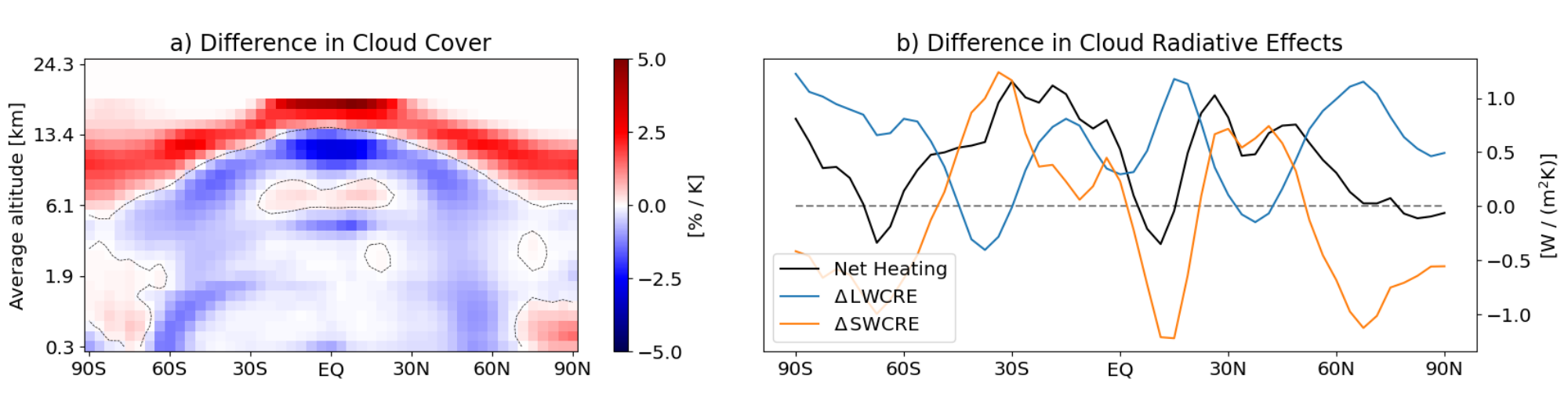}
\caption{Application of our ICON-A-MLe model in a sea surface temperature increase scenario of +4$\,$K: Panel a) illustrates the changes in cloud cover per degree of warming as a function of latitude and altitude relative to a control simulation without induced heating. Dashed contour lines are indicating $0\,\%/\mathrm{K}$. Panel b) presents the corresponding changes in longwave (LW) and shortwave (SW) cloud radiative effects (CRE), and the net cloud-induced heating, i.e., $\Delta \mathrm{LWCRE} + \Delta \mathrm{SWCRE}$, as a function of latitude. Both metrics are computed over the last 10 years of a 20-year simulation (1979–1999) and are normalized by the imposed sea surface temperature increase (4$\,$K)}
\label{fig:iconml_4K}
\end{figure*}

For a climate model to provide reliable future climate projections, it must respond reasonably to changes in climate conditions. To assess the applicability of ICON-A-MLe to a warmer climate, we increase the SST uniformly by 4$\,$K, following a common proxy for global warming scenarios \cite{clark2022correcting, kochkov2024neural, rasp2018deep}. Using the same model setup as in the previous section, we perform a 20-year +4$\,$K simulation with ICON-A-MLe. Figure \ref{fig:iconml_4K} evaluates only the last 10 years of the simulation to allow for model adjustment to the increased SST. \\

The +4$\,$K simulation exhibits a notable rise in the cloud top heights across all latitudes, particularly in the tropics (panel a)). This lifting is consistent with a deepening of the troposphere, a finding supported by observational evidence \cite{Marvel2015, Norris2016}. The depicted decrease of low-level cloud cover at low to mid-latitudes is a direct consequence of increased SSTs, aligning with many current climate model projections \cite{ceppi2017cloud}. This effect is more pronounced at 60°S than at 60°N due to the absence of landmasses in the Southern Hemisphere. Over the Arctic, ICON-A-MLe simulates an increase in low-level clouds under the +4$\,$K scenario, in agreement with observations \cite{Kay2016} and the CESM1 climate model \cite{Morrison2019}. This increase is attributed to sea ice loss during fall. Overall, simulated total cloud cover decreases by 1.65\% in the +4$\,$K scenario. A reduction in total cloud cover with warming is expected, primarily due to the Clausius-Clapeyron relationship \cite{Mendoza2021}. \\
The manually tuned ICON-A baseline model also exhibits these characteristics in the +4$\,$K warming scenario (Fig. \ref{fig_si:fig_S9} a)). While the increase in cloud top altitudes is highly comparable, the decrease of low-level cloudiness at low to mid-latitudes is less pronounced. Furthermore, the increase of low-level Arctic clouds extends to a higher altitude in the troposphere (up to $\approx$5\,km). The latter contributes to a 1.5\% increase in the globally averaged total cloud cover. \\ 
Panel b) presents the difference in CRE between the +4$\,$K and the control simulations, separated into SW and LW components. A positive $\Delta \mathrm{LWCRE}$/$\Delta \mathrm{SWCRE}$ indicates that changes in cloudiness warm the troposphere in the +4$\,$K simulation. The panel shows that the LWCRE of clouds generally increases in the +4$\,$K simulation due to higher and thus colder cloud tops, which emit less LW radiation. Additionally, LWCRE changes are sensitive to the altitude at which cloud cover changes occur. The reduction in low-level cloud cover at 60°S/60°N has very little impact on the warming effect of clouds, whereas decreases in mid-level clouds at 40°S, the Equator, and at 40°N lead to a noticeable reduction in $\Delta$LWCRE. Interpreting the SWCRE differences is more nuanced, as CREs are influenced not only by cloud cover but also by cloud optical properties. Given the changes in panel a), one might expect a symmetric $\Delta \mathrm{SWCRE}$ pattern around the Equator; however, small differences in cloud cover contribute to a significantly reduced albedo effect and increased heating south of the Equator. The observed decrease in the amplitude of SWCRE at 60°S matches the substantial reduction in marine low-level cloud cover. In the Arctic, changes in SWCRE and LWCRE largely offset each other, while in Antarctica, an increase in high-level clouds enhances the positive net CRE. Overall, cloud cover changes in the +4$\,$K simulation contribute to a net increase in tropospheric heating by $0.53\,\mathrm{W}/(\mathrm{m}^2 \mathrm{K})$. \\
In the manually tuned ICON-A baseline model, the change in LWCRE is nearly identical to that of the ICON-A-MLe model (Fig. \ref{fig_si:fig_S9} b)). While generally following a similar zonal pattern, the SWCRE between 70°S and 50°N exhibits a stronger decrease in the +4$\,$K simulation, with a particularly notable reduction of $\approx1\,\mathrm{W}/(\mathrm{m}^2 \mathrm{K})$ between 40°S and 20°N. This enhanced SW cooling is driven by the baseline model's tendency to retain, and partly even increase, low-level tropical cloud cover in the warmer climate (Fig. \ref{fig_si:fig_S9} a)) Collectively, these cloud feedbacks contribute to a net global increase in tropospheric cooling of $0.28\,\mathrm{W}/(\mathrm{m}^2 \mathrm{K})$ in the baseline model. \\

Both the ICON-A-MLe and the baseline ICON-A models project highly similar, physically plausible changes in global precipitation patterns under the +4K warming scenario (Fig. \ref{fig_si:fig_S10}). Specifically, they simulate a strong intensification of rainfall over the central and western Pacific, accompanied by drying over the Maritime Continent and Australia, which resembles El Niño-like conditions. In CMIP3 climate change projections, such conditions are linked to a weakening of the Walker Circulation \cite{vecchi2007global}. The magnitude of the rainfall intensification, averaged over the 10-year period, is up to twice as strong in the ICON-A simulation, peaking at $2.6\cdot 10^{-5}$ kg/(m$^2$sK) in the Andaman Sea. While the monthly mean precipitation distributions for both models exhibit a similar positive shift of approximately $30\%$, ICON-A precipitation is generally stronger by roughly $10^{-4}$ kg/(m$^2$s).

% % Longwave
% - Higher cloud tops increase the greenhouse effect of clouds, as colder clouds emit less LW radiation %
% - LWCRE sensitive to altitude at which cloud cover changes; reduction of low-level cloud cover at 60S/60N does not noticeably decrease cloudiness, while reduction of mid-level cloud cover at 30S/30N does. %
% - Less low cloud cover in extrapolar regions decreases the greenhouse effect slightly (not noticeably due to high air density)
% - More low cloud cover over the Arctic has a stronger effect on the greenhouse than the albedo effect (due to a high solar zenith angle and ice-covered surfaces), counterbalancing the stronger albedo effect due to ...?
% % Shortwave 

% % Generally 
% - Not straightforward to explain panel b) fully by means of panel a). Due to panel a) one might expect a symmetric pattern of $\Delta SWCRE$ in the tropics around the equator, while in practice, minor differences in cloud cover changes cause a significantly lower albedo effect and more heating south of the equator.
% - Overall, we find that in the +4$\,$K simulation changes in cloud coverage further warm the troposphere by $0.43 W/(m^2 K)$

% Concerning Figure 4 (Lohmann)
% swcre = rsut - rsutcs
% lwcre = rlutcs - rlut

% Note: Total cloud cover reduction is 0.06\% for ICON-A.

% SWCRE in ICON-A stronger in tropics thus generally stronger. In ICON-A-ML similar over tropics, weaker over SE pacific, thus weaker overall. 
% Check Rodriguez-Puebla, 2008 whether that is good.
% Both have decreased TOA balance (ICON-A-ML at around -5 W/m^2, ICON-A at around -7 W/m^2).

\section*{Conclusion}
In this study, we present the first application of a machine-learning based parameterization significantly reducing the persistent cloud cover biases in marine stratocumulus regions (by 24.1–54.2\%) and the Southern Ocean (by 75.8\%) in a global climate simulation. We show strong online skill of an offline-trained, data-driven yet physically constrained and interpretable parameterization for cloud cover (equation (\ref{data_driven_eq})) alongside a pipeline for automatically calibrating the resulting hybrid climate model with Earth observations. The data-driven cloud cover parameterization, derived via symbolic distillation from high-resolution data, introduces no additional computational burden and maintains the interpretability of the climate model—distinguishing it from many other machine learning-based approaches. The parameterization is implemented into a climate model (ICON-A) and an automatic tuning procedure is used to optimize the performance of the resulting ICON-A-MLe model against observations. Our tuning approach progressively increases the duration of ICON-A-MLe simulations to keep the required computational resources low. The method is characterized by its simplicity, speed (at the expense of quantifying parameter uncertainty), and extensibility to other parameters and metrics. \\
%The tuning and evaluation of ICON-A-MLe form the primary focus of this study. \\
The tuned ICON-A-MLe model substantially reduces long-standing biases in cloud cover and associated radiative fluxes, particularly in regions dominated by low-level clouds (e.g., the Southern Ocean). More broadly, ICON-A-MLe matches or outperforms the original ICON-A model across a wide range of climate metrics. We use a ``scheme-swap'' experiment to demonstrate that a significant portion of this improvement is directly attributable to the new cloud scheme itself, independent of the tuning procedure. Furthermore, we leverage the scheme's interpretability to identify physical drivers for cloudiness in the Southern Ocean (relative humidity and temperature) and stratocumulus regions (vertical gradient of relative humidity). When subjected to +4 K surface warming, the model captures physically plausible changes in cloud cover, radiative effects, and precipitation—a crucial step towards improving climate projections. \\
While for benchmarking purposes this study focuses on ICON 2.6.4, we anticipate that both the tuning pipeline and the data-driven cloud cover parameterization would yield similar improvements in newer ICON versions or other climate models. Additional enhancements to the tuning procedure could include verifying model performance in different seasons by varying initialization dates before extending to longer simulation periods. Additional metrics and parameters—such as those related to climate system dynamics (e.g., orographic drag or surface wind stress; \cite{giorgetta2018})—could be added to further calibrate the model. Regarding the representation of cloud cover in ICON-A, data-driven equations for cloud inhomogeneity and overlap may be beneficial. \\
We emphasize that our approach—discovering a new data-driven equation that can be used as a parameterization and tuning the hybrid climate model—is not specific to the parameterization of cloud cover. Given an appropriate training set, this pipeline can be applied to any parameterization of ``fast processes'' that one aims to model with a low-dimensional analytical equation. Thus, this work presents a framework for improving climate models using data-driven methods in a practical, computationally efficient, and interpretable way towards hybrid climate models with reduced errors that can provide more robust climate projections for mitigation and adaptation assessments.

\section*{Methods}

\subsection*{Tuning the ICON Climate Model}
In this work we use ICON 2.6.4, a recent version of the ICON climate model with parameterizations still closely based on ECHAM physics (as in \cite{giorgetta2018}). We conduct atmosphere-only simulations at a horizontal resolution of 80\,km with a model time step of 10 minutes, a radiation time step of 2 hours, a vertical grid extending up to 83\,km across 47 levels, initialization on Jan 1, 2016, relabeled to 1979 (as in \cite{giorgetta2018}), and boundary data (e.g., aerosol and ozone concentrations) matching the simulated date. As a first step, we establish a reliable baseline that we can compare to and build upon.

\subsubsection*{Manually Tuned Baseline}
% ICON 1.3.0 ---> ICON 2.6.4
Our default ICON-A 2.6.4 setup operates on a horizontal grid with increased resolution compared to its original description in \cite{giorgetta2018}, and has seen numerous model changes without being recalibrated. As a consequence, a reasonable top-of-the-atmosphere (TOA) radiative balance (difference between incoming and outgoing radiative fluxes) can only be achieved at the cost of overestimating both TOA net incoming shortwave (SW) and outgoing longwave (LW) radiative fluxes (see Fig. \ref{fig_si:fig_S1}). Moreover, total cloud cover is underestimated (global average of 59.3\%), which is too small to be deemed acceptable according to \cite{mauritsen2012tuning}. To address these issues, we also manually tuned ICON-A parameters as an additional reference. Specifically, we varied 13 tuning parameters and utilized approximately 300 simulations from \cite{bonnet2024}, most of them simulating 10 years, to identify which parameters effectively serve as tuning parameters for the radiative metrics and cloud cover (Fig. \ref{fig_si:fig_S2}). Key parameters identified through this process are the coefficient of sedimentation velocity of cloud ice (\textit{cvtfall}), entrainment rate for mid-level convection (\textit{entrmid}), and critical relative humidity for condensation in the upper troposphere (\textit{crt}). By decreasing \textit{cvtfall} from 2.5 to 2.25, allowing ice particles to remain suspended in clouds for longer, slightly increasing \textit{entrmid} from 2e-4 to 2.1e-4 and decreasing \textit{crt} from 0.8 to 0.79, enhancing the formation of high-level clouds at lower relative humidities, we could increase global mean total cloud cover to above 60\% (Fig. \ref{fig_si:fig_S1}). Additionally, these adjustments yielded SW and LW radiative fluxes at the TOA and a radiative balance, both in close agreement with observations. A comprehensive evaluation of the simulation also demonstrates very reasonable zonal means across various metrics (Fig. \ref{fig_si:fig_S3}). By making the convenient yet improper assumption that the impacts of any parameter on a climate metric are linear and independent, we could thus use traditional sensitivity tests to establish a baseline. However, this methodology is both expensive and luck-dependent, highlighting the need for an alternative tuning approach.

\subsubsection*{Automatic Tuning Pipeline} 
We hereby introduce a new automatic tuning pipeline (Fig. \ref{fig:pipeline_sketch}), which we use to tune both the ICON-A baseline model and our ICON-A-MLe model. First, we replace the original cloud cover scheme \cite{sundqvist1989} in ICON-A 2.6.4 by the data-driven equation (equation (\ref{data_driven_eq})) for cloud cover from \cite{grundner2024}, which showed superior skill in a comprehensive offline evaluation. We then define which model parameters are to be adjusted in the tuning process. Their initial values match those obtained from the manually tuned baseline. Specifically, we target 24 different tuning parameters (listed in Table \ref{tab_si:tab_S3}) with a focus on clouds and convection. Ten of the parameters are part of the cloud cover equation and the other 14 are included in the convection (3), cloud optical properties (8), microphysics (2), and vertical diffusion (1) schemes. Our goal is to tune these parameters such that key climate metrics from model simulations are in good agreement with observations. This can be formulated as an optimization problem. Let $p \in \mathbb{R}^{24}$ be the vector with the 24 tuning parameters. Our objective is to find an optimal parameter vector $p_{d}^{\ast}$ that minimizes an objective function, $J_{d}$:
\begin{equation*}
    p_{d}^{\ast} = \arg \min_p J_{d}(p).
\end{equation*}
The objective function $J_{d}$ (formalized as equation (\ref{eq:J}) in the next subsection) quantifies the deviation of a climate model simulation that is based on a specific realization of these parameters from observational reference data. The optimal parameter vector depends on the choice of climate model (implicitly) and on the duration $d$ of the model simulation. As optimization method we use the Nelder-Mead (NM) method (also called `downhill simplex method') \cite{nelder1965simplex}, which is efficient and does not require differentiation of the objective function. To find a minimum, the NM method moves a simplex through the parameter space, iteratively running the ICON model and evaluating the objective function (i.e., the deviation of a model simulation from the reference data) using the parameter values corresponding to each vertex of the simplex in every step. \\

Our final goal is to find a parameter set that minimizes $J_{d}$, where $d$ is on the order of decades. However, to minimize the use of computational resources, we start by simulating only two days. After the NM algorithm has finished tuning the parameters such that the corresponding simulation is as close to the reference data as possible (in practice, when either the objective value is zero or a runtime of 8 hours is reached), we extend the simulation duration to a week, then a month, and finally a year. In each iteration with a given ICON simulation duration, the parameters are initialized using the best setting of the previous iteration. The procedure can be formalized as follows:

\begin{enumerate}

\item First iteration ($d_1 = 2$ days): Find $p_{d_1}^{\ast}$ by minimizing $J_{d_1}(p)$, starting from an initial guess $p_0$.

\item Iterate ($d_i > d_{i-1}$): For subsequent durations ($d_2 = 1$ week, $d_3 = 1$ month, $d_4 = 1$ year), find the new optimum $p_{d_i}^{\ast}$ by minimizing $J_{d_i}(p)$, initializing from the result from the previous stage, $p_{d_{i-1}}^{\ast}$.
\item Minimize $J_{d_4}(p)$ again, now using $P$ from equation (\ref{eq:p_eq}) as the initial guess.
\end{enumerate}
In the last iteration, we avoid multiple full NM cycles using expensive year-long simulations. Instead, we first perform a single NM cycle using a limited number (14--20) of year-long simulations, constrained by the 8-hour runtime limit. This yields a set of candidate parameter vectors, from which we select the one that minimizes the objective function, i.e., $p_{d_4}^{\ast}$, along with its associated TOA energy balance loss, \( T_{d_4} \). In parallel, we retain the best parameter set $p_{d_3}^{\ast}$ from the previous NM tuning iteration, along with its corresponding TOA loss in year-long simulations \( T_{d_3} \). We then construct a new candidate parameter vector \( P \) by extrapolating between \( p_{d_3}^{\ast} \) and \( p_{d_4}^{\ast} \). Specifically, we set
\begin{equation}
\label{eq:p_eq}
P = p_{d_3}^{\ast} - \frac{T_{d_3}}{T_{d_4} - T_{d_3}}(p_{d_4}^{\ast} - p_{d_3}^{\ast}),
\end{equation}
such that the expected TOA loss is minimized under a linear approximation. \\
A second NM cycle is then performed around this extrapolated parameter vector, again using 14--20 year-long simulations. If any of these runs produce a TOA imbalance in the desirable range of 0.5 to 1~W\,m\(^{-2}\), we select the best among them (with the lowest objective value) even if this value is up to 25\% worse than the global minimum. Thereby, we prioritize the TOA balance when sufficiently good solutions are found. Otherwise, we select the parameter set with the lowest objective value. This strategy typically yields high-quality solutions while limiting the number of costly year-long simulations to fewer than 40. \\

\textbf{How do we measure the deviation of a model simulation from given reference data?} First, we define a set of metrics (denoted as M in Fig. \ref{fig:pipeline_sketch}) consisting of globally and temporally averaged (excluding the first day) outgoing TOA LW and SW radiation, the balance between incoming and outgoing radiation, SW and LW cloud radiative effects (CRE), total cloud cover, precipitation, ice and liquid water paths, and water vapor path. Additionally, we consider total cloud cover in the Southeastern Pacific [15–45\textdegree S, 75-105\textdegree W] (averaging at $0.71$), a proxy for representing subtropical stratocumulus clouds, that are typically underestimated in climate models. The LWCRE/SWCRE is defined here as the difference between the TOA outgoing LW/SW radiation in clear-sky conditions and the all-sky (actual) outgoing LW/SW radiation. We then assess whether the simulated metrics fall within the respective ranges provided by multiple observational products (e.g., from CERES \cite{loeb2018clouds}, GPCP-SG \cite{adler2018}, and ISCCP \cite{zhang2023global}). The contribution of each metric to the objective value is quantified as the distance between its computed value and the observational range, normalized by the range width to ensure a relative comparison. \\
Based on the observational products, the expected ranges for outgoing LW and SW radiation are [231, 242] and [100, 110]\,W/m$^2$, respectively, while the net radiation balance should fall within [0.5, 0.9]\,W/m$^2$. Precipitation is expected to range from [2.7, 3.0]\,mm/day, whereas bounds for the remaining metrics are taken from \cite{lauer2023}, Table 5. These bounds are derived from annual means over at least 20 years within the period 1980–2020, and are systematically relaxed (by adding a variability score $\mathrm{varsc}^{max}_{d,m}$ to the upper bound $y^{max}_{obs,m}$ and subtracting $\mathrm{varsc}^{min}_{d,m}$ from the lower bound $y^{min}_{obs,m}$) for shorter simulations to account for natural variability. Using these components, we define the objective value of a model simulation $S_{d}(p)$ of duration $d$ based on parameter values $p$ as 

\begin{equation}
\label{eq:J}
J_{d}(p) = \sum_{m \in \mathrm{M}} \frac{ \max\{S_{d,m}(p) - (y^{max}_{obs, m}+\mathrm{varsc}^{max}_{d,m}),0\} + \max\{y^{min}_{obs,m}-\mathrm{varsc}^{min}_{d,m} - S_{d,m}(p),0\} }{(y^{max}_{obs, m} +  \mathrm{varsc}^{max}_{d,m}) - (y^{min}_{obs,m} - \mathrm{varsc}^{min}_{d,m})}, 
\end{equation}
summing over all metrics $m \in \mathrm{M}$.
This variability scores are estimated by first averaging reference standard ICON-A model outputs over periods matching the current simulation length $d$, and by then taking the differences between a metric's most extreme values and its average value
\begin{align*}
    \mathrm{varsc}^{max}_{d,i} &= \max\{\overline{S_{\mathrm{ref},i}}^d\} - \overline{S_{\mathrm{ref},i}}, \\
    \mathrm{varsc}^{min}_{d,i} &= \overline{S_{\mathrm{ref},i}} - \min\{\overline{S_{\mathrm{ref},i}}^d\}.
\end{align*}
Without such an adjustment of the bounds, shorter simulations would be excessively constrained, limiting the effectiveness of the tuning method. The degree of relaxation varies by metric and simulation length. For example, for global precipitation, the relaxed bounds are [2.34, 3.42], [2.48, 3.23], [2.57, 3.18], and [2.67, 3.03] mm/day for daily, weekly, monthly, and yearly averages, respectively. Note that the required relaxation is relatively small for this metric, as the daily average of global precipitation is already a good approximation of its long-term mean. \\

% CERES, MERRA2, ISCCP, ERA5, GPCP
% https://swift.dkrz.de/v1/dkrz_3a9fa768-cd4d-4c27-b09a-82c2f06a657e/iconeval/ag_atm_amip_r2b5_cvtfall_entrmid_05_cov15_based_on_12962670/index.html

% /home/b/b309170/bd1179_work/bd1179_workspace/calibrating_ml_schemes/objective_bounds.ipynb

% Concerning Figure 2 (pipeline evaluation)
\textbf{How does the tuning pipeline improve the climate model and affect parameter values?} 
At the start of the first tuning iteration, using day-long ICON-A-MLe simulations, cloud cover and cloud water are strongly overestimated, negatively affecting the SW radiative budget and inducing errors in the related tuning metrics (Fig. \ref{fig:pipeline_eval}). These issues are largely resolved halfway through the iteration, most notably through a substantial reduction in the cloud cover offset (by 32.6\%; $a_1$ in equation (equation (\ref{data_driven_eq})) and Table \ref{tab_si:tab_S3}). Towards the end of the iteration, $a_1$ is increased even above its initial value and a remaining underestimation of the water vapor path is corrected through fine-tuning of all parameter values, decreasing the objective value to zero (see also Fig. \ref{fig_si:fig_S4}). The second column of Fig. \ref{fig:pipeline_eval} shows that a 20-year simulation using the parameter values refined in the first iteration already shows substantial improvements in the three radiative metrics, although the radiative balance has not yet been achieved. Such a discrepancy is expected at this stage, as the parameters have only been tuned to produce accurate single-day results so far. \\ 
As the automated tuning process progresses, gradually increasing the ICON-A-MLe simulation duration, the model's skill continues to improve. Eventually, the TOA radiative balance—whose observational estimate lies between 0 and 1\,W/m$^2$—falls within the range of CMIP6 model results. While an objective value of zero is quickly attained in tuning iterations using week- and month-long simulations, in year-long simulations it is decreased from 2.42 to 1.77 (see also Fig. \ref{fig_si:fig_S5}). A final loss of zero is difficult to attain as it requires a full reconciliation of high total cloud cover over the Southeast Pacific with an accurate TOA radiative budget. \\
Table \ref{tab_si:tab_S3} details the parameter modifications due to tuning, with the parameters of the data-driven cloud cover equation (equation (\ref{data_driven_eq})) listed first. Besides the reduction in the cloud cover offset, notable changes include a slight increase in cloud cover sensitivity to relative humidity and temperature, as well as a 42\% decrease in the numerical stabilizer $\epsilon$, which discourages the formation of unrealistic condensate-free clouds. Dependencies on cloud water/ice and vertical relative humidity gradients remain largely unchanged, reinforcing their role in representing thin marine stratocumulus clouds in the model. Adjustments in other parameterizations include setting the Prandtl number slightly above 1 (1.02), modifying cloud droplet number concentrations at low altitudes, and increasing cloud inhomogeneity factors for most cloud types. \\
% Proof of plausibility concerning the Prandtl number: https://journals.ametsoc.org/view/journals/atsc/72/6/jas-d-14-0335.1.xml?tab_body=pdf, Fig. 6
In general, we find that it is beneficial to modify many parameters slightly, which is difficult to achieve manually. We find that those parameters that undergo significant changes already do so in the first tuning iteration with day-long simulations (e.g., \texttt{cn2sea} is changed from $80$ to $41.71$). In subsequent iterations, parameter adjustments typically become smaller and tend to stabilize. Exceptions include the numerical stabilizer, which is further reduced in the month-long iteration (from $0.80$ to $0.62$). \\
The overall computational cost of tuning the ICON-A-MLe model automatically amounts to approximately 400 node-hours (using 12 AMD EPYC 7763 MILAN CPU nodes in parallel), running 741/29/32/28 ICON-A-MLe simulations in the tuning iterations using daily/weekly/monthly/yearly simulations, respectively. \\

\textbf{Automatic tuning of the ICON-A baseline.}
To ensure maximum comparability between the ICON-A-MLe and ICON-A models, we apply the same automatic tuning approach to the ICON-A baseline. For comparison, tuning the ICON-A baseline does not yield any single dominant parameter modification. Some parameters that were decreased in the ICON-A-MLe tuning (e.g., mid-level entrainment rate, autoconversion of cloud droplets to rain) are instead increased (see Table \ref{tab_si:tab_S4}), suggesting a dependency on the cloud cover scheme and demonstrating Nelder-Mead’s ability to adjust parameters accordingly. A notable modification is an increase of the critical relative humidity at the top (by $3.4$\%) in the iteration with weekly simulations, which reduces the simulated amount of high clouds, correcting an issue of overestimated outgoing SW radiation (reducing it by $1.8\, \mathrm{W}/\mathrm{m}^2$). In the case of the ICON-A baseline we found that by nudging the critical relative humidities for cloud formation (from 0.99 to 0.968 at the top and from 0.59 to 0.665 at the bottom of the atmosphere) before the second NM cycle, we end up with a slightly better ICON-A model configuration (TOA loss of zero and an objective value of 3.52 instead of 3.78 without the adjustment). \\
Throughout the tuning process, the challenge of balancing high total cloud cover over the Southeast Pacific with an accurate TOA radiative budget remains more evident in the traditional atmospheric model. Therefore, the total computational cost is higher, amounting to approximately 600 node-hours, with 432/327/158/36 ICON-A simulations in the four tuning iterations.

\subsection*{Data-Driven Cloud Cover Equation}
The data-driven equation for cloud cover from \cite{grundner2024}, discovered with symbolic regression \cite{cranmer2023interpretable} on coarse-grained data from storm-resolving DYAMOND simulations \cite{stevens2019dyamond, stephan2022atmospheric}, can be formulated as follows:
\begin{equation}
\label{data_driven_eq}
    f(\mathrm{RH}, \mathrm{T}, \partial_z\mathrm{RH}, \mathrm{q}_c, \mathrm{q}_i) = \mathrm{I}_1(\mathrm{RH}, \mathrm{T}) + \mathrm{I}_2(\partial_z\mathrm{RH}) + \mathrm{I}_3(\mathrm{q}_c, \mathrm{q}_i),
\end{equation}
a function of relative humidity ($\mathrm{RH}$), its vertical derivative ($\partial_z\mathrm{RH}$), temperature (T), cloud water ($\mathrm{q}_c$), cloud ice content ($\mathrm{q}_i$) with 
\begin{align*}
    \mathrm{I}_1(\mathrm{RH}, \mathrm{T}) &= a_1 + a_2(\mathrm{RH} - \overline{\mathrm{RH}}) + a_3(\mathrm{T} - \overline{\mathrm{T}})\\ &+ \frac{a_4}{2} (\mathrm{RH}- \overline{\mathrm{RH}})^2
    + \frac{a_5}{2}(\mathrm{T} - \overline{\mathrm{T}})^2(\mathrm{RH}- \overline{\mathrm{RH}}) \\
    \mathrm{I}_2(\partial_z\mathrm{RH}) &= a_6^3\left(\partial_z\mathrm{RH} + \frac{3a_7}{2}\right)\left(\partial_z\mathrm{RH}\right)^2 \\
    \mathrm{I}_3(\mathrm{q}_c, \mathrm{q}_i) &= \frac{-1}{\mathrm{q}_c/a_8 + \mathrm{q}_i/a_9 + \epsilon}.
\end{align*}
After automatically tuning equation (\ref{data_driven_eq}) online within ICON-A, its attained parameter values are 
\begin{align*}
    \{a_1, \dotsc, a_9, \epsilon\} = \{0.118, 1.234, -0.0265\,\mathrm{K}^{-1}, 5.65, 1.56\cdot10^{-3}\,\mathrm{K}^{-2},& \\
    591.68\,\mathrm{m}, 2.22\,\mathrm{km}^{-1}, 1.47\,\mathrm{mg/kg}, 0.344\,\mathrm{mg/kg}, 0.615\}&    
\end{align*}
(see also Table \ref{tab_si:tab_S3}). The values for the reference relative humidity $\overline{\mathrm{RH}}$ ($0.6025$) and temperature $\overline{\mathrm{T}}$ ($257.06\,\mathrm{K}$) are fixed.

\subsection*{Implementation of the Data-Driven Scheme in ICON-A}
The data-driven cloud cover parameterization is implemented as a diagnostic, three-dimensional cloud fraction scheme within the $\text{ICON-A}$ model (version 2.6.4) \cite{giorgetta2018}. It is computed at every model time step for each atmospheric grid cell, using the local thermodynamic state (e.g., $\text{T}$, $\text{RH}$, and their vertical gradients) from that cell as inputs for equation (\ref{data_driven_eq}). This scheme replaces the model's native Sundqvist-based diagnostic cloud fraction calculation \cite{sundqvist1989}. It interacts with other model components in the following ways:
\begin{enumerate}
\item Cloud Microphysics: The model's existing one-moment microphysics scheme continues to prognose the liquid and ice water mixing ratios ($q_l$, $q_i$) \cite{lohmann1996}. Our new scheme provides the cloud fraction ($\mathcal{C}$) required to compute the tendencies of liquid and ice water and water vapor in each grid cell.
\item Radiation: These 3D fields ($\mathcal{C}$ and grid-mean condensates) are passed to the $\text{PSrad}$ radiation scheme \cite{pincus2013}. The radiation scheme's internal vertical overlap logic (a standard maximum-random assumption) is unchanged. $\text{PSrad}$ operates on these 3D inputs to compute the final 2D radiative fluxes at the top of the atmosphere and at the surface.
\end{enumerate}

\section*{Data availability}
Algorithms, code, and data for reproducing our results are available at our GitHub page \url{https://github.com/EyringMLClimateGroup/grundner25_iconaml_automatic_tuning/tree/main} and preserved at \url{https://doi.org/10.5281/zenodo.15194060}.

% \section*{Code availability}
% Mention the availability of custom code or software used.

% \printbibliography

\section*{Funding}
A.G., J.S., M.S., and V.E. received funding for this study from the European Research Council (ERC) Synergy Grant ``Understanding and modeling the Earth System with Machine Learning (USMILE)'' under the Horizon 2020 research and innovation programme (Grant agreement No. 855187) and from the Horizon Europe project ``Artificial Intelligence for enhanced representation of processes and extremes in Earth System Models (AI4PEX)'' (Grant agreement ID: 101137682). T.B. received support from AIPEX, funded by the Swiss State Secretariat for Education, Research and Innovation (SERI, Grant No. 23.00546). V.E. was additionally supported by the Deutsche Forschungsgemeinschaft (DFG, German Research Foundation) through the Gottfried Wilhelm Leibniz Prize awarded to V.E. (reference no. EY 22/2-1).

\section*{Acknowledgements}
This work used resources of the Deutsches Klimarechenzentrum (DKRZ) granted by its Scientific Steering Committee (WLA) under project ID bd1179.

\section*{Author contributions}
A.G., T.B. and V.E. designed the research; A.G. and J.S. performed the research and analyzed the data; A.L. and M.S. created the concept and ESMValTool code used for the evaluation of the simulations; A.G. wrote the paper with contributions from all co-authors.

\section*{Competing interests}
The authors declare no conflict of interest.

%\bibliography{bibfile}

\newpage
\clearpage

% Begin of Supplementary Information

\setcounter{figure}{0}

\renewcommand{\thefigure}{S\arabic{figure}}
\renewcommand{\thetable}{S\arabic{table}}

\begin{center}
  \LARGE \textbf{Supplementary Information}\par\vspace{0.5em}
  {\large \textit{Reduced cloud cover errors in a hybrid AI-climate model through equation discovery and automatic tuning}}\par\vspace{1em}
  \vspace{1em}
  \small Arthur Grundner, Tom Beucler, Julien Savre, Axel Lauer, Manuel Schlund, and Veronika Eyring
\end{center}

\newpage

\section*{Supplementary Figures}

\begin{figure}[tbhp]
\centering
\hspace*{-5em}
% \captionsetup{margin=2cm} % To confine the caption to the image
\includegraphics[width=1.3\textwidth]{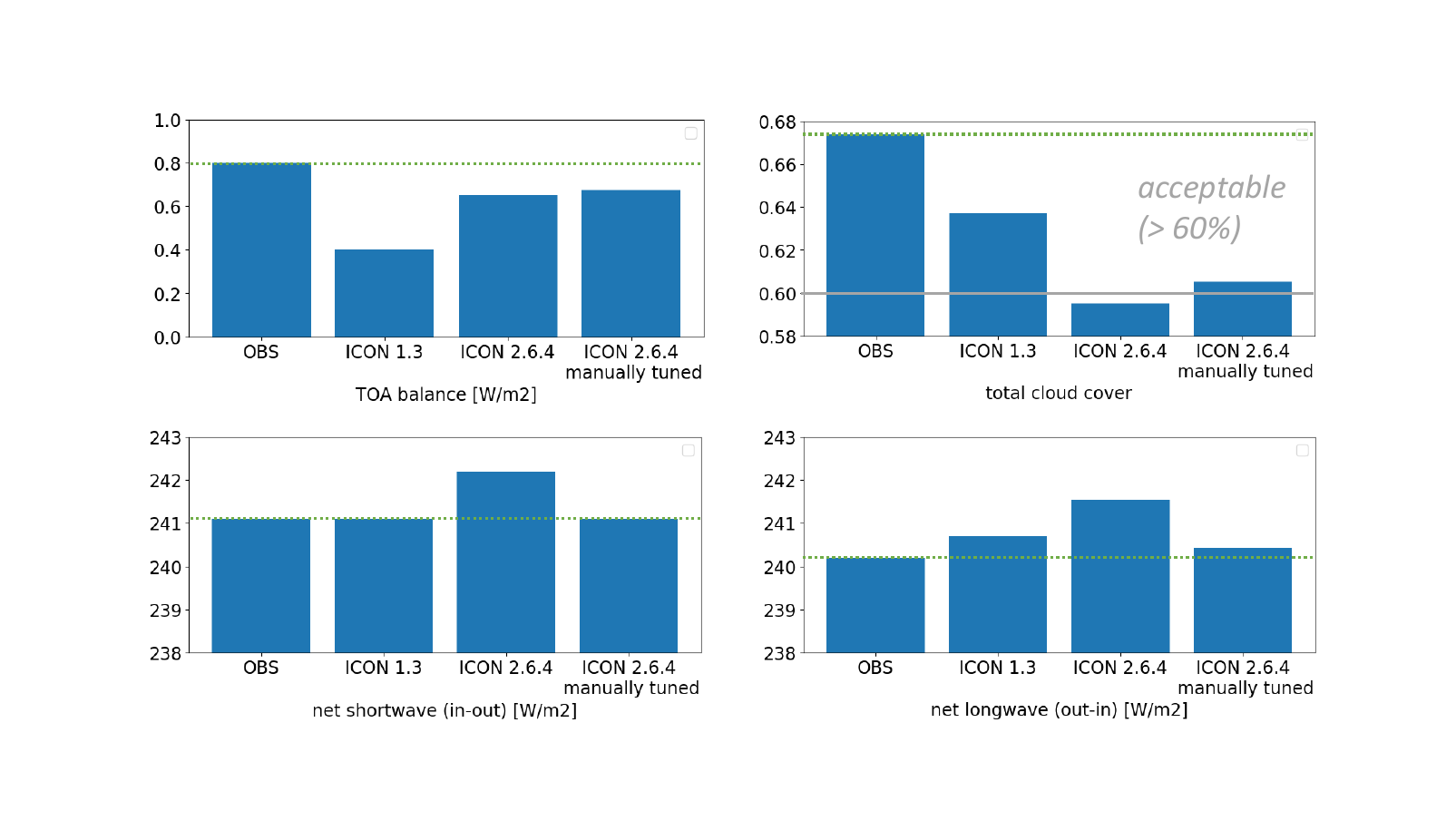}
\caption{Selection of globally and temporally averaged climate metrics of three 10-year ICON simulations (1980-1989). The ICON-A 2.6.4. simulations used an 80\,km horizontal grid, while the original ICON-A 1.3 model simulation ran at an 160\,km resolution. The dashed green lines highlight the values from observations, taken from \cite{giorgetta2018}. The lower bound for acceptable total cloud cover values is due to \cite{mauritsen2012tuning}.}
\label{fig_si:fig_S1}
\end{figure}

\begin{figure}[tbhp]
\centering
\hspace*{-4cm}
% \captionsetup{margin=1cm} % To confine the caption to the image
\includegraphics[width=1.4\textwidth]{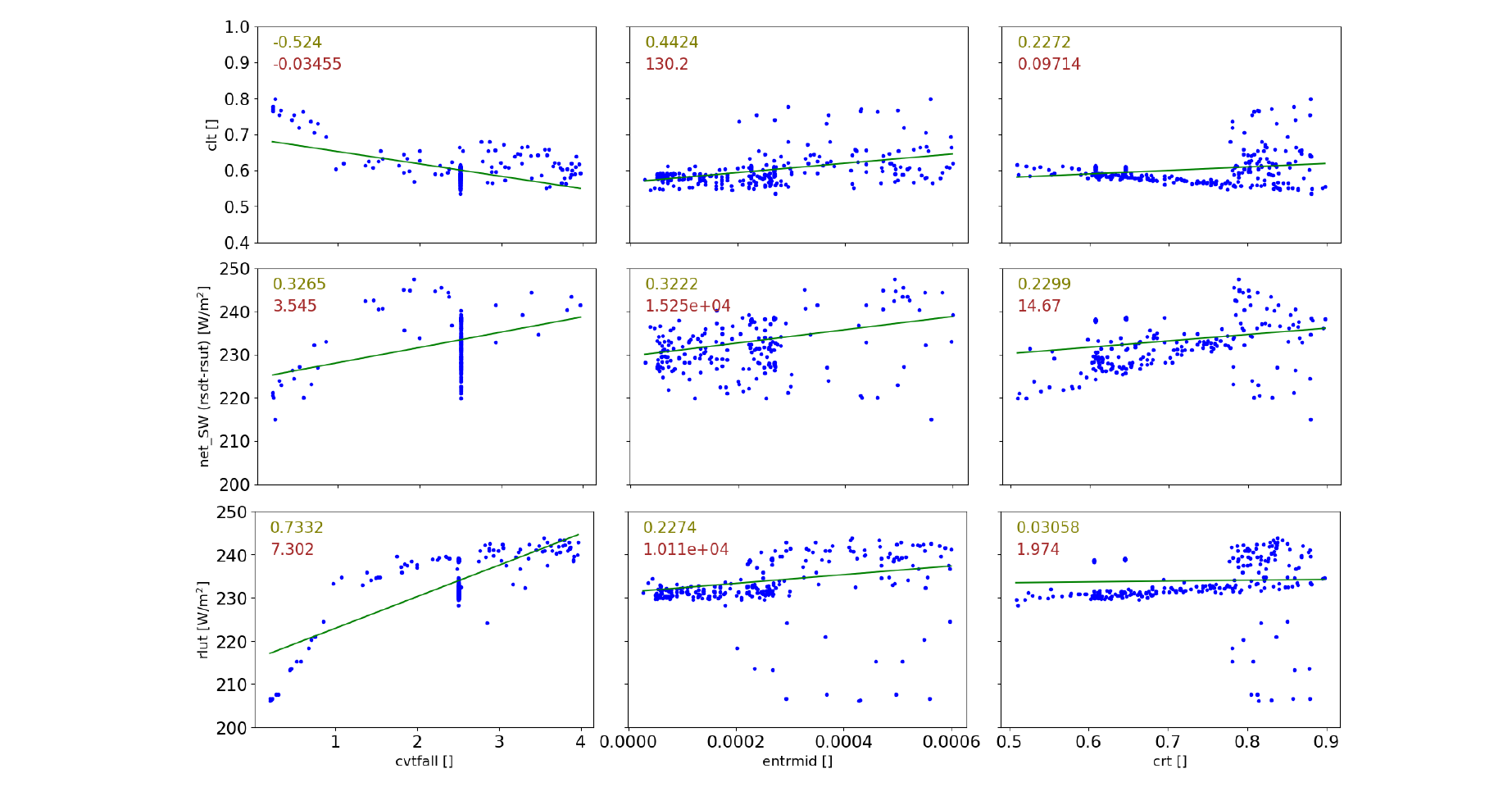}
\caption{297 ICON-A simulations with a set of 13 different perturbed parameters (of which only the effects of three are shown). The ordinates describe the global and temporal averages of total cloud cover (clt), the net shortwave and upwelling longwave radiation (rlut) at the top of the atmosphere. The first number reported in each panel are the Pearson correlation coefficients between a given parameter and metric, indicating that \textit{cvtfall}, \textit{entrmid} and \textit{crt} are generally correlated with these three metrics. The second number is the slope of the linear regression line, suggesting how much a parameter should be adjusted to achieve a desired change in a given metric.}
\label{fig_si:fig_S2}
\end{figure}

\begin{figure}[tbhp]
\centering
% \captionsetup{margin=1cm} % To confine the caption to the image
\hspace*{-4.5cm}
\includegraphics[width=1.5\textwidth]{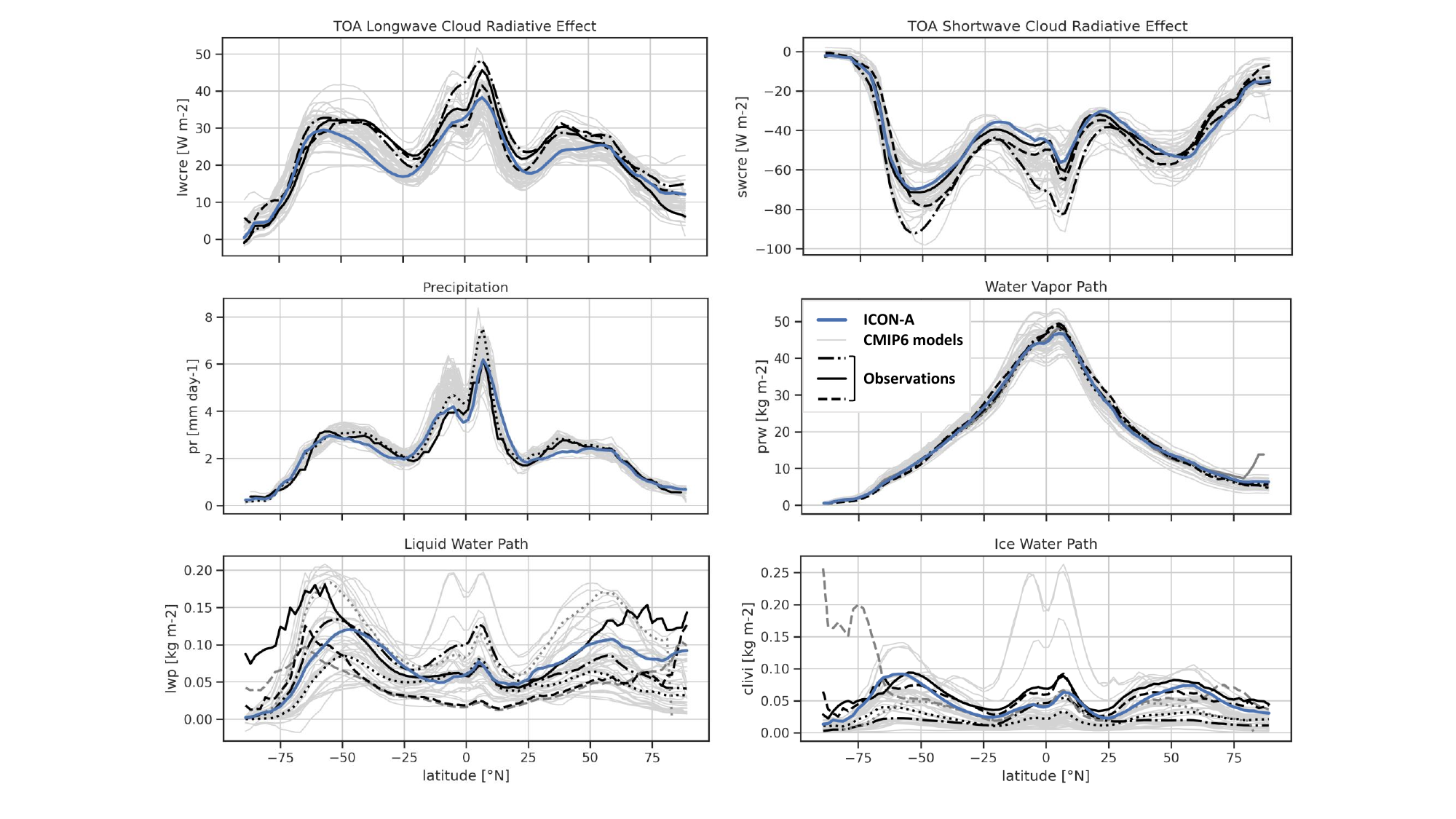}
\caption{Zonal means of nine important climate variables. The blue lines show zonal means from a 20-year simulation (1979-1999) with the manually tuned ICON-A model, the solid gray lines are based on established CMIP6 model simulations \cite{eyring2016overview} and all other lines correspond to observational sources and reanalysis products.}
\label{fig_si:fig_S3}
\end{figure}

\begin{figure}
\centering
\captionsetup{margin=1cm} % To confine the caption to the image
\hspace*{-7.5cm}
\includegraphics[width=1.9\textwidth]{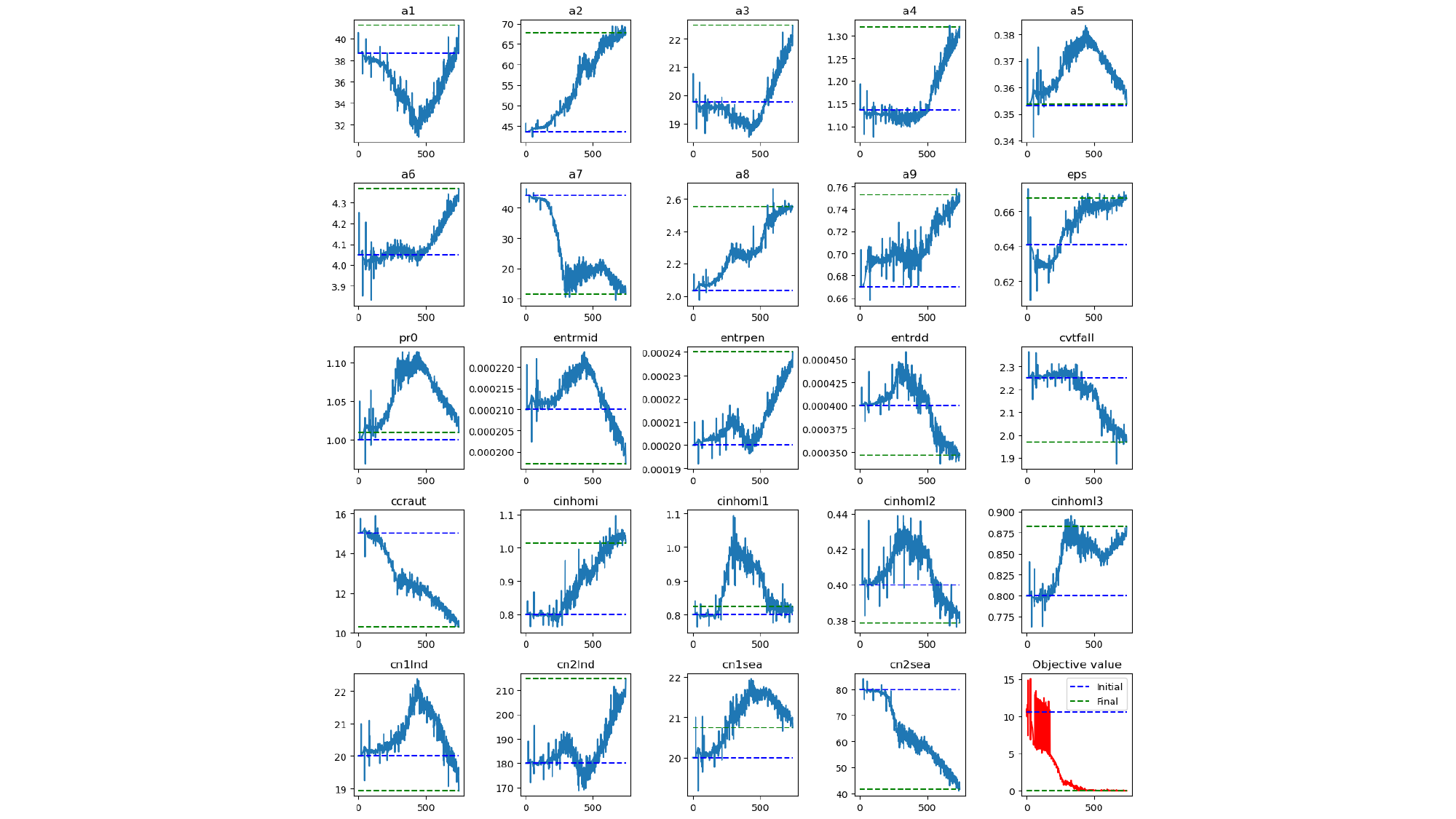}
\caption{Evolution of tuning parameters (for a description see Table \ref{tab_si:tab_S3}) while tuning the ICON-A-MLe model using day-long simulations. The dashed blue line indicates the initial and the dashed green line the final selected parameter values.}
\label{fig_si:fig_S4}
\end{figure}

\begin{figure}
\centering
\captionsetup{margin=1cm} % To confine the caption to the image
\hspace*{-4cm}
\includegraphics[width=1.4\textwidth]{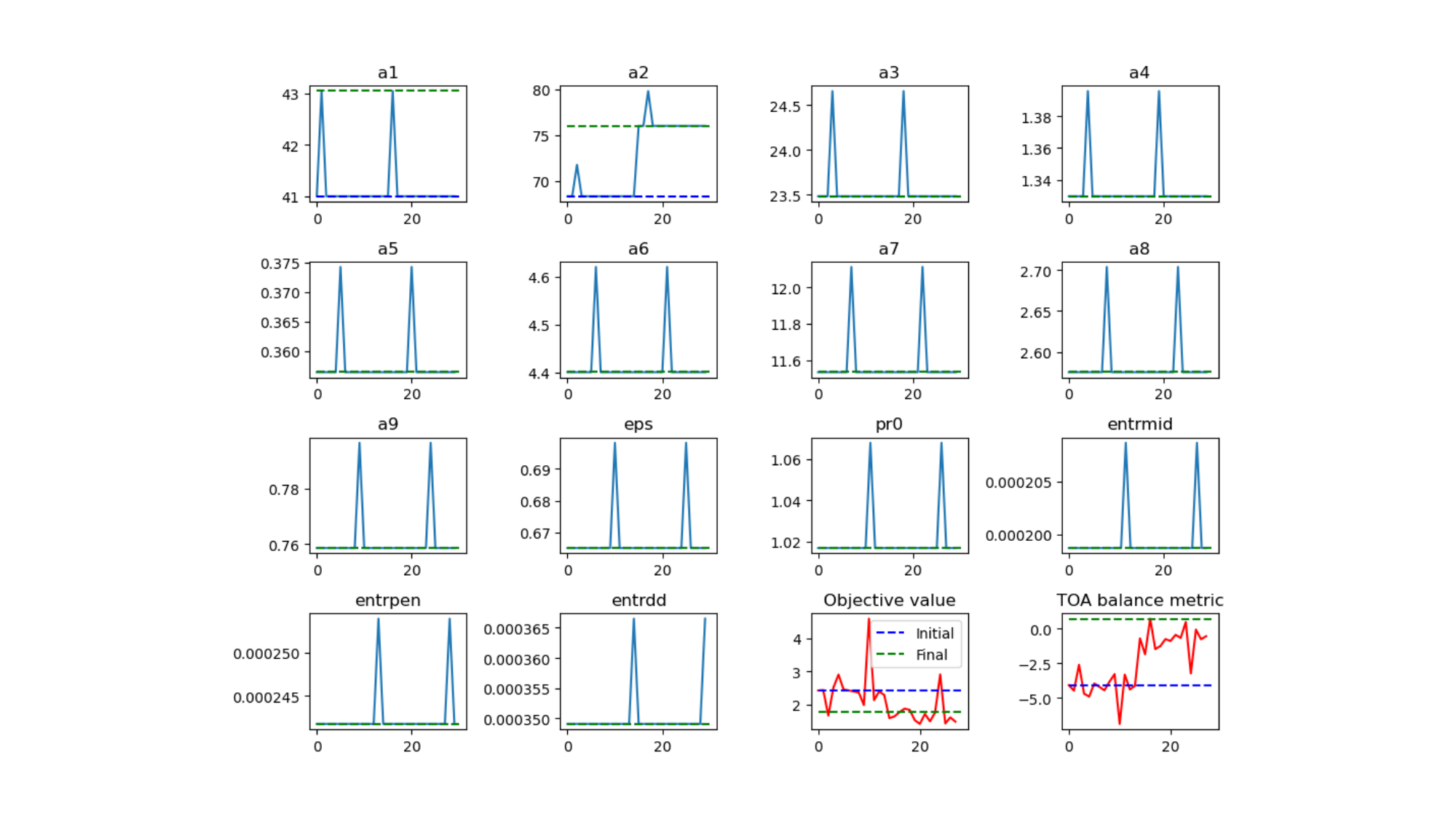}
\caption{Evolution of tuning parameters (for a description see Table \ref{tab_si:tab_S3}) while tuning the ICON-A-MLe model using year-long simulations. The dashed blue line indicates the initial and the dashed green line the final selected parameter values. The parameters that are not shown remain constant.}
\label{fig_si:fig_S5}
\end{figure}

\begin{figure}[tbhp]
\centering
\hspace*{-2cm}
\includegraphics[width=1.2\textwidth]{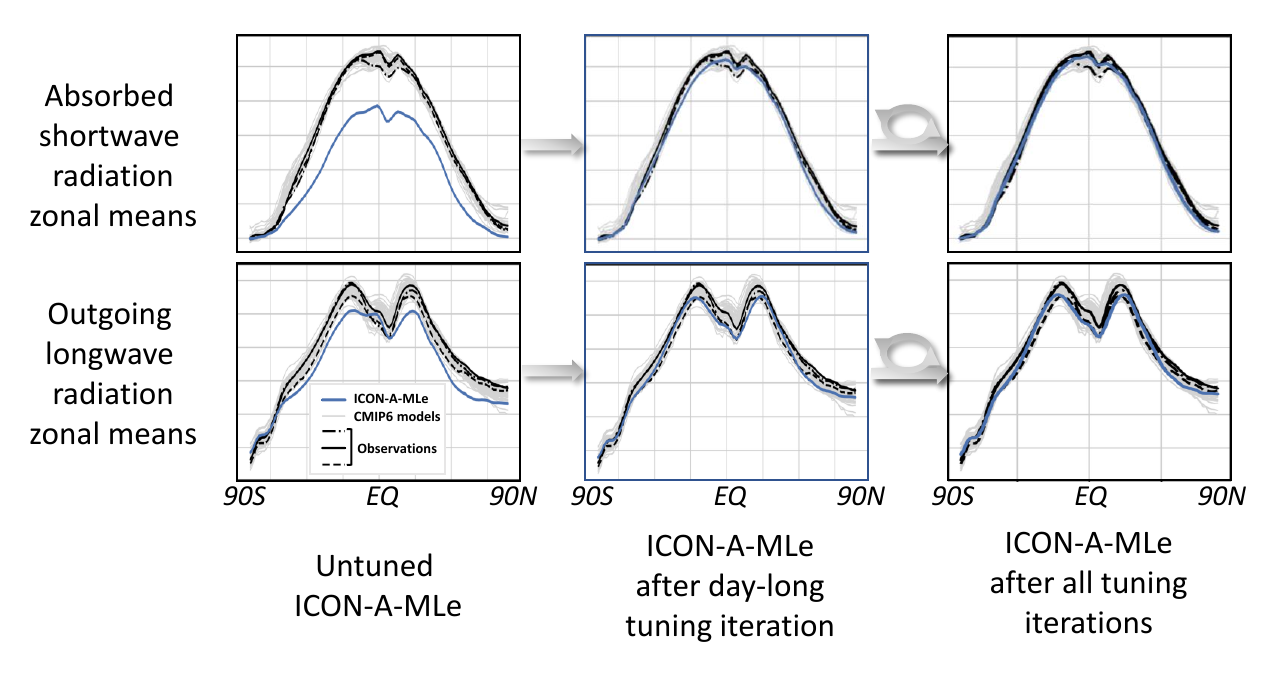}
\caption{Like Fig. \ref{fig:pipeline_eval}, but showing zonal means of the top of the atmosphere longwave and shortwave radiative fluxes.}
\label{fig_si:fig_S6}
\end{figure}

\begin{figure}[tbhp]
\centering
\hspace*{-2.4cm}
% \captionsetup{margin=1.6cm} % To confine the caption to the image
\includegraphics[width=1.3\textwidth]{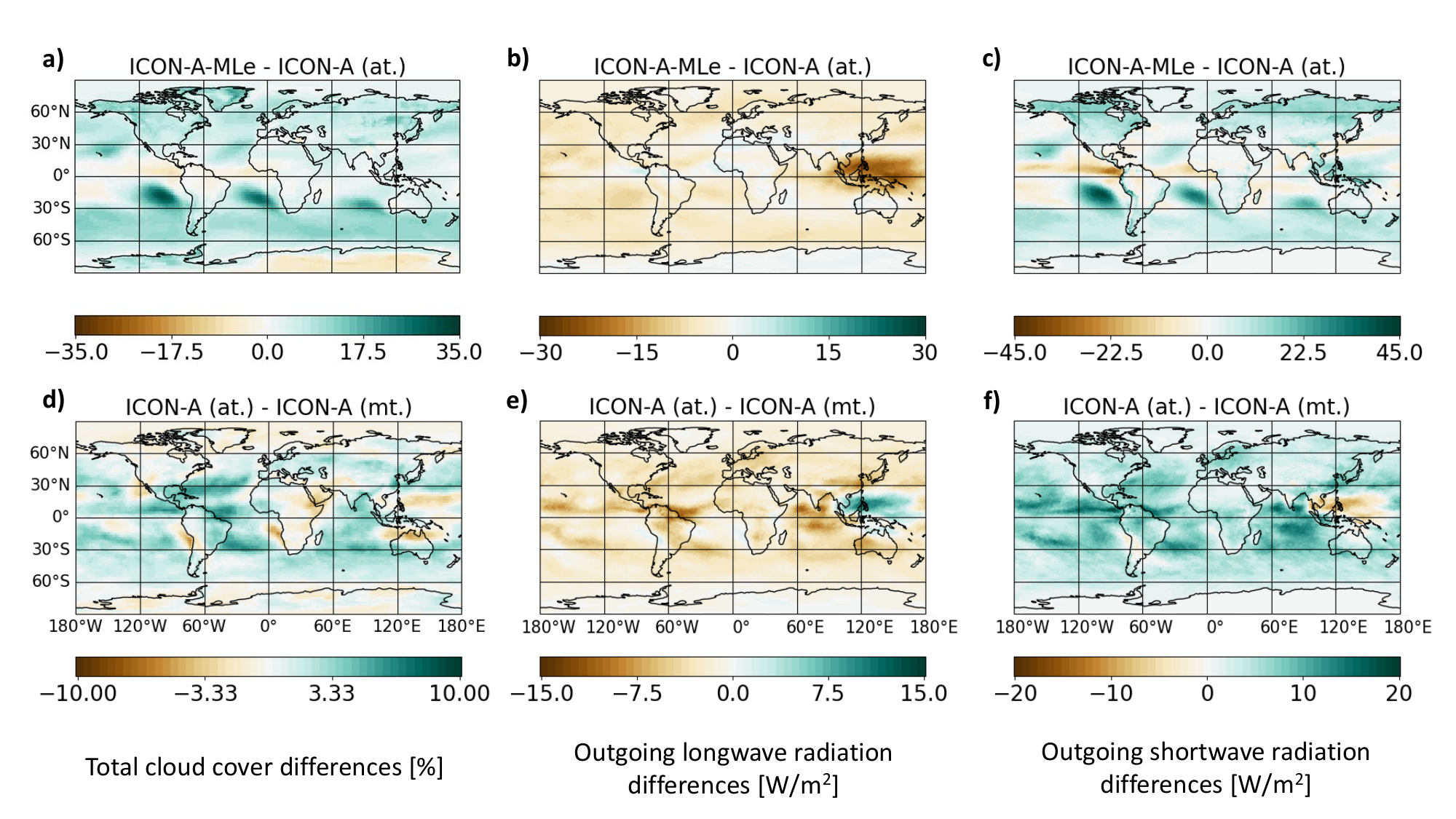}
\caption{Differences between the panels of each column in Fig. \ref{fig:iconml_eval}, showing bias differences among the automatically tuned (at.) ICON-A-MLe, ICON-A, and manually tuned (mt.) ICON-A models for three key climate metrics over the 1979–1999 period.}
\label{fig_si:fig_S7}
\end{figure}

\begin{figure}[tbhp]
\centering
\hspace*{-5cm}
% \captionsetup{margin=1.6cm} % To confine the caption to the image
\includegraphics[width=1.6\textwidth]{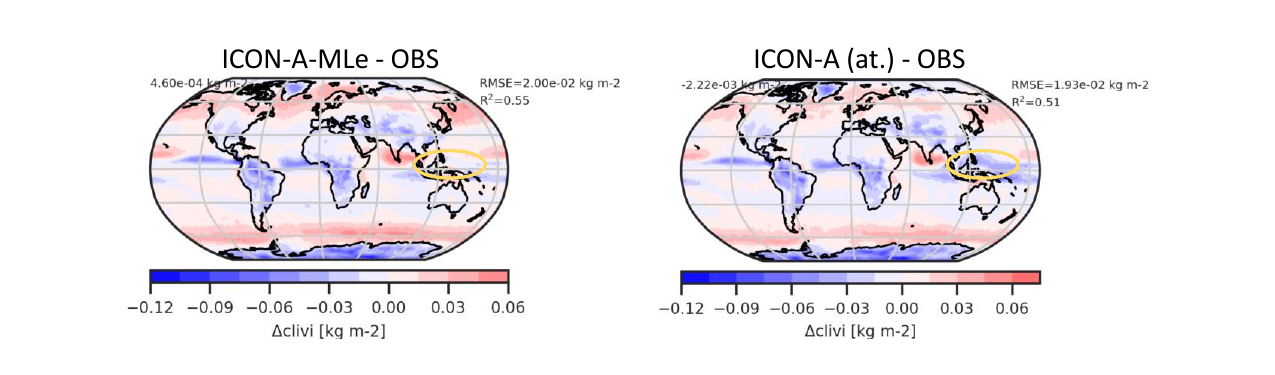}
\caption{Like Fig. \ref{fig:iconml_eval}, but showing column-integrated cloud ice (ice water path) for the ICON-A-MLe and the automatically tuned ICON-A model simulations. The region highlighted in yellow corresponds to the region around the Philippines in which the ICON-A-MLe model has an increased outgoing longwave radiation bias. The number in the top left provides the difference between the simulated and observational global averages.}
\label{fig_si:fig_S8}
\end{figure}

\begin{figure}[tbhp]
\centering
\hspace*{-1.7cm}
% \captionsetup{margin=1.6cm} % To confine the caption to the image
\includegraphics[width=1.2\textwidth]{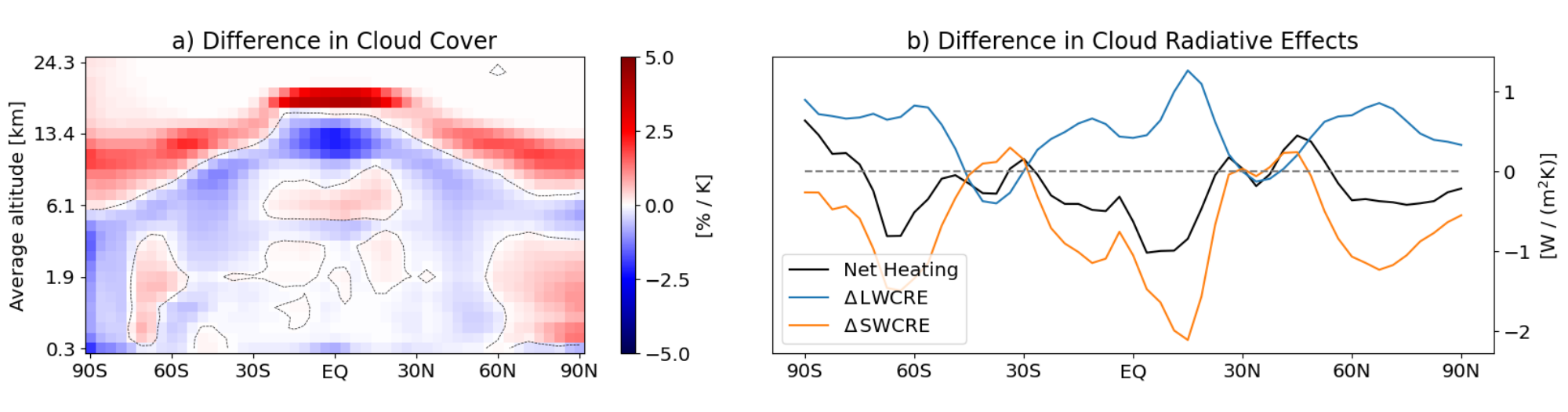}
\caption{Like Fig. \ref{fig:iconml_4K}, but using the manually tuned ICON-A baseline model to conduct the control and +4K warming scenario simulations.}
\label{fig_si:fig_S9}
\end{figure}

\begin{figure}[tbhp]
\centering
\hspace*{-6cm}
% \captionsetup{margin=1.6cm} % To confine the caption to the image
\includegraphics[width=1.7\textwidth]{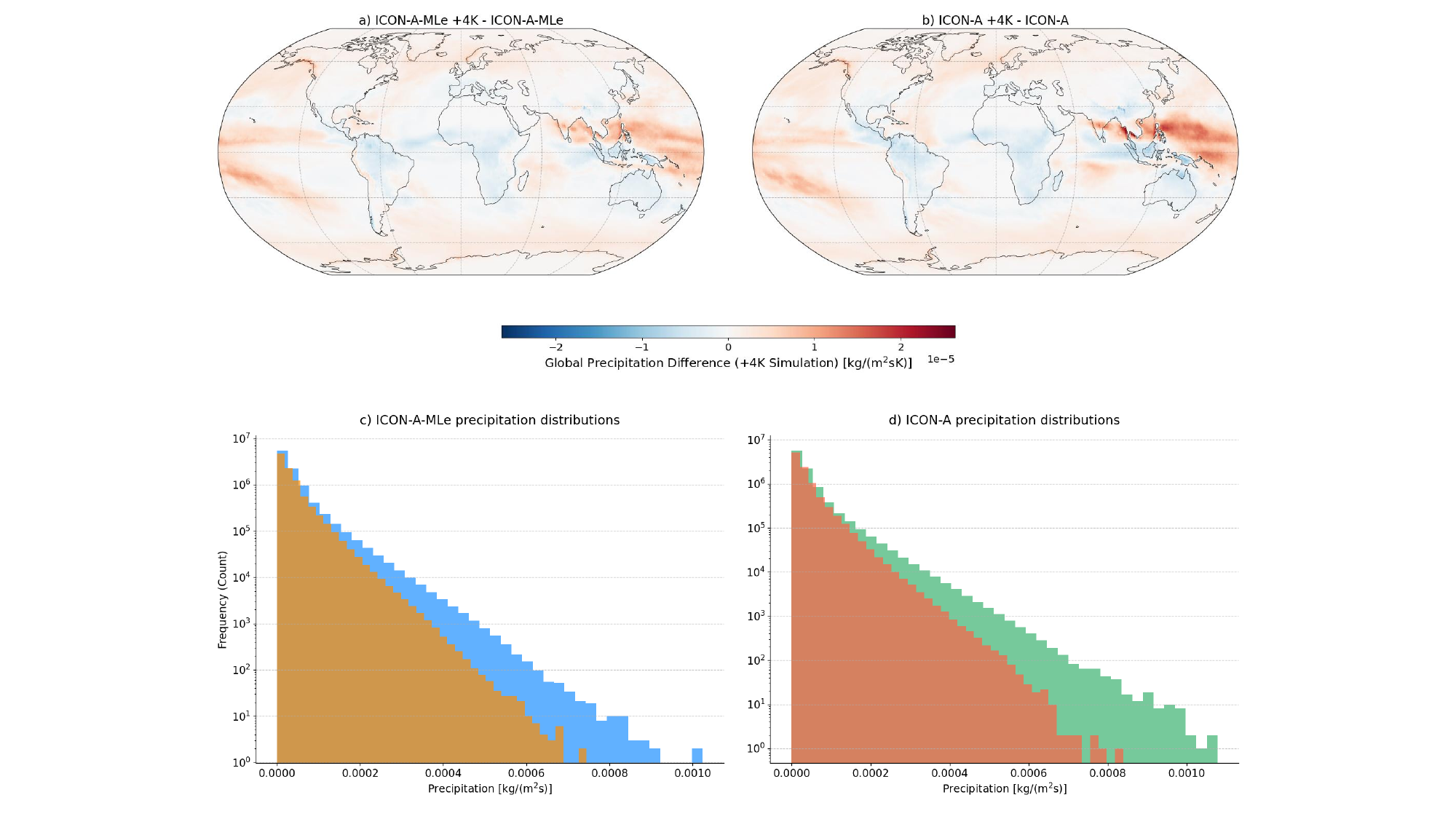}
\caption{Precipitation metrics. Panels a) and b) illustrate the changes in precipitation per degree of warming relative to a corresponding ICON-A-MLe (at.)/ICON-A (mt.) control simulation without induced heating. Panels c) and d) show histograms of monthly averaged precipitation values. The values are taken from the last 10 years of a 20-year simulation (1979-1999)}
\label{fig_si:fig_S10}
\end{figure}

\clearpage

\section*{Supplementary Tables}

\begin{table}[tbhp]
\centering
\small
% \captionsetup{margin=2.2cm} % To confine the caption to the image
\caption{Evaluation of time-averaged two-dimensional climate variables against observations for our ICON-A-MLe and the native ICON-A simulations. The lower the root mean-squared error (RMSE) and the higher the coefficient of determination (R$^2$-value) the better. The lowest RMSE values for a given metric across the simulations of Table \ref{tab_si:tab_S1} and \ref{tab_si:tab_S2} are highlighted in bold.}
\begin{tabular}{lrrrr}
& \multicolumn{3}{r}{RMSE (R$^2$-values)} \\
\cmidrule(lr){2-4} 
Metric & ICON-A-MLe & ICON-A (at.) & ICON-A (mt.) & Reference \\
\midrule
Near-Surface Air Temperature & 1.22\,K (0.99) & 1.23\,K (0.99) & 1.2\,K (0.99) & ERA5 \\
Precipitation & 1.05\,mm/day (0.74) & 1.04\,mm/day (0.74) & 1.09\,mm/day (0.73) & GPCP-SG \\
Ice Water Path & 2e-2\,kg/m2 (0.55) & 1.93e-2\,kg/m2 (0.51) & \textbf{1.85e-2\,kg/m2} (0.57) & MultiOBS1\\
Liquid Water Path & 2.73e-2\,kg/m2 (0.56) & 2.96e-2\,kg/m2 (0.34) & 3.38e-2\,kg/m2 (0.45) & MultiOBS1 \\
Condensed Water Path & 3.34e-2\,kg/m2 (0.71) & 3.33e-2\,kg/m2 (0.67) & 3.87e-2\,kg/m2 (0.66) & MultiOBS1 \\
Water Vapor Path & 2.19\,kg/m2 (0.98) & 2.06\,kg/m2 (0.98) & 2.03\,kg/m2 (0.98) & MultiOBS2 \\
LW cloud radiative effect & 7.22\,W/m2 (0.71) & \textbf{4.37\,W/m2} (0.85) & 8.65\,W/m2 (0.39) & MultiOBS3 \\
SW cloud radiative effect & \textbf{10.19\,W/m2} (0.79) & 10.29\,W/m2 (0.77) & 11.43\,W/m2 (0.72) & MultiOBS3 \\
Total Cloud Cover & 9.87\% (0.68) & 11.66\% (0.58) & 12.64\% (0.52) & MultiOBS4 \\
\bottomrule
\multicolumn{4}{l}{\textit{MultiOBS1 := \{CLARA-AVHRR, CLOUDSAT-L2, ERA5, ESACCI-CLOUD, MERRA2, MODIS\}}} \\
\multicolumn{4}{l}{\textit{MultiOBS2 := \{ERA5, ESACCI-WATERVAPOUR, ISCCP-FH, MERRA2\}}} \\
\multicolumn{4}{l}{\textit{MultiOBS3 := \{CERES-EBAF, ESACCI-CLOUD, ISCCP-FH, MERRA2\}}}  \\
\multicolumn{4}{l}{\textit{MultiOBS4 := \{CLARA-AVHRR, ERA5, ESACCI-CLOUD, MERRA2, MODIS, PATMOS-x\}}} 
\end{tabular}
\label{tab_si:tab_S1}
\end{table}

\begin{table}[tbhp]
\centering
\small
% \captionsetup{margin=2.2cm} % To confine the caption to the image
\caption{Biases of ICON with the data-driven scheme in the auto-tuned ICON-A configuration (ICON-A-MLe$^*$) and of ICON with the native scheme in the auto-tuned ICON-A-MLe configuration (ICON-A$^*$) with subsequent auto-tuning of the cloud cover scheme parameters. The lower the root mean-squared error (RMSE) and the higher the coefficient of determination (R$^2$-value) the better. The column $\text{ML} - \neg \text{ML}$ lists ((ICON-A-MLe$^*$ $-$ ICON-A) $+$ (ICON-A-MLe $-$ ICON-A$^*$))/2, a measure of how much the data-driven scheme improves a metric.}
\begin{tabular}{lrrrr}
& \multicolumn{3}{r}{RMSE (R$^2$-values)} \\
\cmidrule(lr){2-4} 
Metric & ICON-A-MLe$^*$ & ICON-A$^*$ & $\text{ML} - \neg \text{ML}$ & Reference \\
\midrule
Near-Surf. Air Temp. & 1.22\,K (0.99) & \textbf{1.13\,K} (0.99) & 0.04\,K (0.0) & ERA5 \\
Precipitation & \textbf{0.98\,mm/day} (0.77) & 1.12\,mm/day (0.7) & -0.07\,mm/day (0.04) & GPCP-SG \\
Ice Water Path & 1.88e-2\,kg/m2 (0.56) & 2.04e-2\,kg/m2 (0.5) & -0.06e-2\,kg/m2 (0.05) & MultiOBS1 \\
Liquid Water Path & \textbf{2.05e-2\,kg/m2} (0.73) & 4.21e-2\,kg/m2 (-0.05) & -1.18e-2\,kg/m2 (0.5) & MultiOBS1 \\
Condensed Water Path & \textbf{3.15e-2\,kg/m2} (0.73) & 4.16e-2\,kg/m2 (0.52) & -0.51e-2\,kg/m2 (0.13) & MultiOBS1 \\
Water Vapor Path & \textbf{2.00\,kg/m2} (0.98) & 2.01\,kg/m2 (0.98) & 0.01\,kg/m2 (0.0) & MultiOBS2 \\
LW cloud radiative effect & 5.42\,W/m2 (0.81) & 6.25\,W/m2 (0.75) & 1.01\,W/m2 (-0.09) & MultiOBS3 \\
SW cloud radiative effect & 11.78\,W/m2 (0.73) & 11.54\,W/m2 (0.75) & 0.07\,W/m2 (0.0) & MultiOBS3 \\
Total Cloud Cover & \textbf{9.71\%} (0.69) & 11.59\% (0.54) & -1.77\% (0.13) & MultiOBS4 \\
% lwcre & 5.36\,W/m2 (0.8) & 6.33\,W/m2 (0.75) & XX & XX \\
% swcre & 11.86\,W/m2 (0.72) & 11.56\,W/m2 (0.75) & XX & XX \\
\bottomrule
\end{tabular}
\label{tab_si:tab_S2}
\end{table}

\begin{table}[tbhp]
\centering
% \captionsetup{margin=1.5cm} % To confine the caption to the image
\small
\caption{Selected ICON-A-MLe(*) tuning parameters and their values. The first 10 listed parameters are part of the data-driven cloud cover equation described in \cite{grundner2024}.}
\hspace*{-5em}
\begin{tabular}{rrrrr}
& & \multicolumn{2}{r}{Tuned values} \\
\cmidrule(lr){3-4}
Parameter & Default values & ICON-A-MLe & ICON-A-MLe* & Short description \\
\midrule
% These are from the *latest* provided param(:) values for ICON-A-MLe*
\texttt{a1} & $0.444$ & $0.118$ & $0.248$ & Cloud cover offset \\ % param(1)
\texttt{a2} & $1.164$ & $1.234$ & $0.778$ & Relative humidity sensitivity coefficient \\ % param(2)
\texttt{a3} & $-0.015$ & $-0.027$ & $-0.023$ & Temperature sensitivity coefficient [1/K] \\ % param(3)
\texttt{a4} & $4.067$ & $5.65$ & $3.741$ & Quadratic relative humidity sensitivity coefficient \\ % param(4)
\texttt{a5} & $1.32 \cdot 10^{-3}$ & $1.56 \cdot 10^{-3}$ & $1.132 \cdot 10^{-3}$ & Temperature-relative humidity interaction coefficient [1/K$^2$] \\ % param(5)
\texttt{a6} & $590.01$ & $591.68$ & $421.724$ & Vertical relative humidity gradient sensitivity scale [m] \\ % param(6)
\texttt{a7} & $2.07 \cdot 10^{-3}$ & $2.22 \cdot 10^{-3}$ & $5.133 \cdot 10^{-3}$ & Vertical relative humidity gradient offset coefficient [1/m] \\ % param(7)
\texttt{a8} & $1.16 \cdot 10^{-6}$ & $1.47 \cdot 10^{-6}$ & $1.831 \cdot 10^{-6}$ & Liquid condensate scaling coefficient [kg/kg] \\ % param(8)
\texttt{a9} & $3.07 \cdot 10^{-7}$ & $3.44 \cdot 10^{-7}$ & $3.582 \cdot 10^{-7}$ & Ice condensate scaling coefficient [kg/kg] \\ % param(9)
\texttt{eps} & $1.06$ & $0.615$ & $0.749$ & Small positive numerical stabilizer \\ % param(10)

% Other parameters follow, using the values from the list you provided two turns ago for ICON-A-MLe*
\texttt{pr0} & $1$ & $1.017$ & $1.059$ & Neutral limit Prandtl number \\
\texttt{entrmid} & $2.1 \cdot 10^{-4}$ & $1.99 \cdot 10^{-4}$ & $2.18 \cdot 10^{-4}$ & Entrainment rate for midlevel convection \\
\texttt{entrpen} & $2.0 \cdot 10^{-4}$ & $2.42 \cdot 10^{-4}$ & $2.06 \cdot 10^{-4}$ & Entrainment rate for penetrative convection \\
\texttt{entrdd} & $4.0 \cdot 10^{-4}$ & $3.49 \cdot 10^{-4}$ & $3.98 \cdot 10^{-4}$ & Entrainment rate for cumulus downdrafts \\
\texttt{cvtfall} & $2.25$ & $1.984$ & $2.186$ & Coefficient of sedimentation velocity of cloud ice \\
\texttt{ccraut} & $15$ & $10.359$ & $16.697$ & Coefficient of autoconversion of cloud droplets to rain \\
\texttt{cinhomi} & $0.8$ & $0.915$ & $0.846$ & Ice cloud inhomogeneity factor \\
\texttt{cinhoml1} & $0.8$ & $0.830$ & $0.865$ & Liquid cloud inhomogeneity factor, stratiform clouds \\
\texttt{cinhoml2} & $0.4$ & $0.382$ & $0.390$ & Liquid cloud inhomogeneity factor, shallow convection \\
\texttt{cinhoml3} & $0.8$ & $0.889$ & $0.832$ & Liquid cloud inhomogeneity factor, several types of convection \\
\texttt{cn1lnd} & $20$ & $19.084$ & $20.987$ & Cloud droplet num. conc. over land, high altitude [1/cm$^3$] \\
\texttt{cn2lnd} & $180$ & $216.285$ & $177.610$ & Cloud droplet num. conc. over land, low altitude [1/cm$^3$] \\
\texttt{cn1sea} & $20$ & $20.901$ & $20.206$ & Cloud droplet num. conc. over sea, high altitude [1/cm$^3$] \\
\texttt{cn2sea} & $80$ & $42.038$ & $81.950$ & Cloud droplet num. conc. over sea, low altitude [1/cm$^3$] \\
\bottomrule
\end{tabular}
\label{tab_si:tab_S3}
\end{table}

\begin{table}[tbhp]
\centering
% \captionsetup{margin=1.5cm} % To confine the caption to the image
\small
\caption{Selected ICON-A(*) tuning parameters and their values. The first five parameters (\textit{crs, crt, csatsc, nex, cinv}) belong to the native ICON-A cloud cover scheme.}
\hspace*{-5em}
\begin{tabular}{rrrrr}
& & \multicolumn{2}{r}{Tuned values} \\
\cmidrule(lr){3-4}
Parameter & Default values & ICON-A & ICON-A* & Short description \\
\midrule
\texttt{crs} & $0.968$ & $0.968$ & $0.990$ & Critical relative humidity at the surface \\
\texttt{crt} & $0.8$ & $0.698$ & $0.577$ & Critical relative humidity at the top \\
\texttt{csatsc} & $0.7$ & $0.702$ & $0.703$ & Minimum saturation for cloud cover below marine inversion \\
\texttt{nex} & $2$ & $2.169$ & $2.211$ & Transition parameter for critical relative humidity profile \\
\texttt{cinv} & $0.25$ & $0.249$ & $0.272$ & Fraction of dry adiabatic lapse rate for inversion top over sea \\
\texttt{pr0} & $1$ & $1.059$ & $1.017$ & Neutral limit Prandtl number \\
\texttt{entrmid} & $2.1 \cdot 10^{-4}$ & $2.18 \cdot 10^{-4}$ & $1.99 \cdot 10^{-4}$ & Entrainment rate for midlevel convection \\
\texttt{entrpen} & $2.0 \cdot 10^{-4}$ & $2.06 \cdot 10^{-4}$ & $2.42 \cdot 10^{-4}$ & Entrainment rate for penetrative convection \\
\texttt{entrdd} & $4.0 \cdot 10^{-4}$ & $3.98 \cdot 10^{-4}$ & $3.49 \cdot 10^{-4}$ & Entrainment rate for cumulus downdrafts \\
\texttt{cvtfall} & $2.25$ & $2.186$ & $1.984$ & Coefficient of sedimentation velocity of cloud ice \\
\texttt{ccraut} & $15$ & $16.697$ & $10.359$ & Coefficient of autoconversion of cloud droplets to rain \\
\texttt{cinhomi} & $0.8$ & $0.846$ & $0.915$ & Ice cloud inhomogeneity factor \\
\texttt{cinhoml1} & $0.8$ & $0.865$ & $0.830$ & Liquid cloud inhomogeneity factor, stratiform clouds \\
\texttt{cinhoml2} & $0.4$ & $0.390$ & $0.382$ & Liquid cloud inhomogeneity factor, shallow convection \\
\texttt{cinhoml3} & $0.8$ & $0.832$ & $0.889$ & Liquid cloud inhomogeneity factor, several types of convection \\
\texttt{cn1lnd} & $20$ & $20.987$ & $19.084$ & Cloud droplet number concentration over land, high altitude [1/cm$^3$] \\
\texttt{cn2lnd} & $180$ & $177.610$ & $216.285$ & Cloud droplet number concentration over land, low altitude [1/cm$^3$] \\
\texttt{cn1sea} & $20$ & $20.206$ & $20.901$ & Cloud droplet number concentration over sea, high altitude [1/cm$^3$] \\
\texttt{cn2sea} & $80$ & $81.950$ & $42.038$ & Cloud droplet number concentration over sea, low altitude [1/cm$^3$] \\
\bottomrule
\end{tabular}
\label{tab_si:tab_S4}
\end{table}

\newpage
\clearpage

\bibliography{bibfile}

\end{document}